# Anisotropic galvanomagnetic effects in single-crystal Fe(001) films elucidated by a phenomenological theory

Haoran Chen[1], Zhen Cheng[1], Yizi Feng[1], Hongyue Xu[1], Tong Wu[1], Chuanhang Chen[1], Yue Chen[2,3,4], Zhe Yuan[3,4], Yizheng Wu[1,5,6*]

[1]*Department of Physics, Applied Surface Physics State Key Laboratory, Fudan University, Shanghai 200433, China*

[2]*Center for Advanced Quantum Studies and Department of Physics, Beijing Normal University, Beijing 100875, China*

[3]*Institute for Nanoelectronic Devices and Quantum Computing, Fudan University, Shanghai 200433, China*

[4]*Interdisciplinary Center for Theoretical Physics and Information Science, Fudan University, Shanghai 200433, China*

[5]*Shanghai Research Center for Quantum Sciences, Shanghai 201315, China*

[6]*Shanghai Key Laboratory of Metasurfaces for Light Manipulation, Fudan University, Shanghai 200433, China*

## Abstract

Utilizing the phenomenological theory based on crystal symmetry operation, we have established the complete angular dependencies of the galvanomagnetic effects, encompassing both anisotropic magnetoresistance (AMR) and the planar Hall effect (PHE), for the ferromagnetic films with $C_{4v}$ symmetry. These dependencies were experimentally confirmed via comprehensive angular-mapping of AMR and PHE in single-crystal Fe(001) films at room temperature. We demonstrated that the intrinsic magnetization-induced effects are independent of the field strength by carefully separating the field-induced and magnetization-induced galvanomagnetic effects. Our theoretical and experimental findings highlight the absence of in-plane four-fold angular dependence in PHE, a feature prohibited by the Onsager relation in systems with $C_4$ symmetry. This study affirms that the universal angular dependencies of AMR and PHE in single crystals can be accurately predicted by the conventional phenomenological theory.

# I. Introduction

Anisotropic magnetoresistance (AMR), a fundamental spin-dependent transport phenomenon, was first observed by Lord Kelvin in 1856 [1]. AMR describes the relationship between longitudinal electrical resistivity ($\rho_{xx}$) and the orientation of magnetization ($\mathbf{m}$) in ferromagnetic materials. The phenomenon of AMR has garnered significant interest, serving as the foundation for numerous technological advancements, including magnetic field sensors and memory devices [2-4], which are essential components of contemporary electronics and data storage systems.

Generally, the mechanism of AMR can be understood as the anisotropic s-d scattering of electrons induced by spin-orbit coupling (SOC) [5-8]. This mechanism gives rise to a two-fold $\cos 2\alpha$ signal in $\rho_{xx}$, where $\alpha$ denotes the angle between the magnetization $\mathbf{m}$ and the electric current $\mathbf{J}$. In magnetic films, when $\mathbf{m}$ rotates in the film plane, the transverse resistivity $\rho_{xy}$ is observed to change with the magnetization orientation with the relation $\sin 2\alpha$, which is known as planar Hall effect (PHE) [9-11]. The PHE is often regarded as the transverse counterpart of AMR, due to their similar physical origins. Both AMR and PHE are categorized as galvanomagnetic effects.

Studies of galvanomagnetic effects have been extensively conducted in single crystals, where the angular dependence of AMR is related to crystal structure. The angular-dependent AMR of single crystals was first systematically investigated in Ni single crystals [12-14]. Such pioneering work was followed by studies on single-crystal Fe [14,15,16] and Co [17]. To elucidate the connection between AMR and crystal symmetry, a phenomenological theory was formulated [12]. This theory successfully provides an expression for $\rho_{xx}$ as a function of the magnetization orientation. The phenomenological theory of galvanomagnetic effects was subsequently refined by Birss [18] and Potter [5], enabling the derivation of the angular-dependence of $\rho_{xy}$ as well. The phenomenological theory reveals a reciprocal relationship in (001) films of cubic crystals between AMR for $\mathbf{J} \parallel [100]$ and PHE for $\mathbf{J} \parallel [110]$, as well as between AMR for $\mathbf{J} \parallel [110]$ and PHE for $\mathbf{J} \parallel [100]$, which has been validated experimentally [19,20].

However, the phenomenological theory, while providing expressions for AMR and PHE based on symmetry operation, falls short of elucidating the physical origins of each angular-dependent term. Early theoretical studies [5-8] predominantly attributed AMR to the scattering of itinerant s-band electrons by localized d-band electrons. Utilizing first-principle electronic band calculations, recent studies on single-crystal CoFe [19] and FePt [21] films have revealed a connection between AMR and the intricate electronic band structure. The underlying connection of AMR with electronic band structure in other single-crystal materials remains an active area of investigation.

In most experimental studies on angular-dependent AMR [17,22-28], the magnetization is rotated within the $xy$, $yz$, and $xz$ plane, with the direction of the electric current $\mathbf{J}$ defined as the $x$ axis, the transverse direction as the $y$ axis, and the out-of-plane direction as the $z$ axis. The results of these field-rotation measurements are consistent with the phenomenological model. However, the capability of the phenomenological theory to describe galvanomagnetic effects with arbitrarily oriented $\mathbf{m}$ has seldom been extensively examined in experiments. An additional issue that has often been overlooked in literatures is the identification and separation of magnetic-field-induced magnetoresistance. Although the field-induced magnetoresistance (FIMR) due to the Lorentz force effect has been well recognized in the field-sweeping measurements, it was usually ignored while discussing the magnetization-induced AMR by varying magnetic field orientations. The magnetoresistance measured in experiments is the addition of field-induced effect and magnetization-induced effect. FIMR is also anisotropic and has contribution to the total signal in angular-dependent measurements.

Different from the conventional phenomenological theory initialized by Döring [12], a recent phenomenological theory adopted a new strategy to decompose the resistivity tensor into angular-dependent terms [20] and provided the angular dependence of $\rho_{xx}$ and $\rho_{xy}$ when magnetization rotates in the $xy$, $yz$, and $xz$ planes in cubic crystals. This new theory has been reported to agree with the experimental measurements of various current orientations [20]. Although the conventional phenomenological theory has been well applied in literatures to describe the AMR in single crystal systems, most experiments were performed with the field applied only in the $xy$ plane [15,16,21,26,29]. One question still remains whether the phenomenological theory is sufficient to describe the magnetization-induced AMR in single crystals. Thus, a systematic experimental study of full angular-mapping of magnetoresistance with arbitrary magnetization orientation [30], followed by a comprehensive comparison with phenomenological theory, is required.

In this study, we applied the conventional phenomenological theory to derive the full angular dependence of galvanomagnetic effects in systems with $C_{4v}$ symmetry, considering arbitrary orientations of magnetization and current. To validate this theory, we performed comprehensive angular-mapping measurements of AMR and PHE in single crystal Fe(001) films at room temperature (RT). Our experimental fitting revealed that only eight parameters are sufficient to describe the measured full set of longitudinal and transverse resistivity curves. The reciprocal relationship between AMR and PHE with current along [100] and [110] directions was confirmed through both experimental observations and theoretical analysis. Furthermore, our results indicate that the intrinsic magnetization-induced galvanomagnetic effects are independent of the field strength, following a meticulous determination of the angular relationship between magnetization and magnetic field, and a separation of the effects induced by magnetization from those induced by the external field. Collectively, our findings demonstrate that the conventional phenomenological theory adequately captures the angular dependencies of AMR and PHE in Fe(001) films.

## II. Phenomenological theory of galvanomagnetic effects

The phenomenological theory of galvanomagnetic effects has been widely applied to understand the angular dependence of AMR in single crystals [5,6,12,17-19,31]. In the following, we describe the phenomenological theory of galvanomagnetic effects, and derive $\rho_{xx}$ and $\rho_{xy}$ and their dependences on magnetization orientation and current direction in single-crystal magnetic films with cubic lattice structure.

As described by Ohm's law, the relation between electric field $\mathbf{E}$ and electric current $\mathbf{J}$ can be expressed as $E_i = \rho_{ij}J_j$, and the subscripts $i$ and $j$ take the value of 1, 2, and 3 with Einstein summation, corresponding to orthogonal coordinate axes. The resistivity tensor $\rho_{ij}$ is a function of the unit magnetization vector $\mathbf{m}$, and is expanded in a MacLaurin's series [5]

$$\rho_{ij}(\mathbf{m}) = a_{ij} + a_{kij}m_k + a_{klij}m_k m_l + \cdots, \tag{1}$$

where the Einstein summation convention is used. Here, we only focus on the resistivity symmetric to magnetization with the even-ordered expansion terms contributing to AMR, and the anomalous Hall effect (AHE) related to the odd-ordered terms [18] are not discussed in this study. Thus, if the expansion is truncated up to the 4th order, the resistivity tensor $\rho_{ij}$ for AMR and PHE can be expressed as

$$\rho_{ij}(\mathbf{m}) = a_{ij} + a_{klij}m_k m_l + a_{klmnij}m_k m_l m_m m_n. \tag{2}$$

Due to Onsager reciprocal relation $\rho_{ij}(\mathbf{m}) = \rho_{ji}(-\mathbf{m})$ and Neumann's principle [5], the total number of free parameters $a_{ij}$, $a_{klij}$ and $a_{klmnij}$ can be greatly reduced. Bulk Fe crystal has $O_h$ symmetry [5], but in the Fe(001) films, the $C_{4v}$ symmetry should be considered [31]. Through the symmetry transformation, the number of independent parameters in $\rho_{ij}(\mathbf{m})$ can be reduced to be 15.

For the Fe(001) film, the current can only be applied in the (001) plane, thus the resistivity tensor's components $\rho_{11}$, $\rho_{12}$, $\rho_{21}$, and $\rho_{22}$ can have contribution to the measured longitudinal and Hall voltages. If defining $\theta_{\mathbf{mc}}$ and $\phi_{\mathbf{mc}}$ as the polar angle and in-plane azimuthal angle of $\mathbf{m}$in respective to the crystal axes, we can express the magnetization unit $\mathbf{m}$ in spherical coordinate system with $\mathbf{m} = (\sin\theta_{\mathbf{mc}}\cos\phi_{\mathbf{mc}}, \sin\theta_{\mathbf{mc}}\sin\phi_{\mathbf{mc}}, \cos\theta_{\mathbf{mc}})$. After the symmetry analysis, we can obtain the following components of resistivity tensor:

$$\begin{aligned} \rho_{11}(\theta_{\mathbf{mc}}, \phi_{\mathbf{mc}}) &= A_1 + A_2\cos2\phi_{\mathbf{mc}} + A_4\cos4\phi_{\mathbf{mc}}, \\ \rho_{22}(\theta_{\mathbf{mc}}, \phi_{\mathbf{mc}}) &= A_1 - A_2\cos2\phi_{\mathbf{mc}} + A_4\cos4\phi_{\mathbf{mc}}, \\ \rho_{12}(\theta_{\mathbf{mc}}, \phi_{\mathbf{mc}}) &= \rho_{21}(\theta_{\mathbf{mc}}, \phi_{\mathbf{mc}}) = A_3\sin2\phi_{\mathbf{mc}}. \end{aligned} \tag{3}$$

$A_1$, $A_2$, $A_3$ and $A_4$ are the $\theta_{\mathbf{mc}}$-dependent coefficients, which can be expressed as

$$\begin{aligned} A_1(\theta_{\mathbf{mc}}) &= \rho_0 + \rho_1\cos2\theta_{\mathbf{mc}} + \rho_2\cos4\theta_{\mathbf{mc}}, \\ A_2(\theta_{\mathbf{mc}}) &= \rho_3(1 - \cos2\theta_{\mathbf{mc}}) - \rho_5(1 - \cos4\theta_{\mathbf{mc}}), \\ A_3(\theta_{\mathbf{mc}}) &= \rho_4(1 - \cos2\theta_{\mathbf{mc}}) - \rho_6(1 - \cos4\theta_{\mathbf{mc}}), \\ A_4(\theta_{\mathbf{mc}}) &= \rho_7(3 - 4\cos2\theta_{\mathbf{mc}} + \cos4\theta_{\mathbf{mc}}), \end{aligned} \tag{4}$$

There are only 8 free parameters $\rho_0, \rho_1, \cdots, \rho_7$, which are the combination of the original parameters $a_{ij}$, $a_{klij}$, $a_{klmnij}$ [32]. In bulk Fe(001) with $O_h$ symmetry, the number of free parameters in $\rho_{ij}$ is further reduced to 6, consistent with the literature [5,18,19].

Next, the longitudinal resistivity $\rho_{xx}^{\mathbf{m}}$ and planar Hall resistivity $\rho_{xy}^{\mathbf{m}}$ are derived, with current applied in the (001) plane and arbitrary magnetization directions. In $\rho_{xx}^{\mathbf{m}}$ and $\rho_{xy}^{\mathbf{m}}$, the superscript $\mathbf{m}$ is added to indicate the resistivity changed by the magnetization orientation. Our study only emphasizes the intrinsic galvanomagnetic effects induced by the change of magnetization orientation, thus the magnetic-field-induced changes of $\rho_{xx}$ and $\rho_{xy}$ are first neglected in the phenomenological theory, but will be discussed in the following experimental studies. The current direction can be described by $\hat{l} = (\cos\phi_{\mathrm{J}}, \sin\phi_{\mathrm{J}}, 0)$ with $\phi_{\mathrm{J}}$ as the angle between the current and the [100] axis, and the transversal vector for Hall measurement is $\hat{t} = (-\sin\phi_{\mathrm{J}}, \cos\phi_{\mathrm{J}}, 0)$. In most experimental studies on AMR [17,22-28], the angular-dependent AMR is expressed in respective to the angle between magnetization and current. So, we further define the current direction as the $x$ direction. In this coordination, the magnetization unit $\mathbf{m}$ can also be expressed by the polar angle $\theta_{\mathbf{m}}$ and in-plane azimuthal angle $\phi_{\mathbf{m}}$. With the relation $\theta_{\mathbf{mc}} = \theta_{\mathbf{m}}$ and $\phi_{\mathbf{mc}} = \phi_{\mathbf{m}} + \phi_{\mathrm{J}}$, both $\rho_{xx}^{\mathbf{m}}$ and $\rho_{xy}^{\mathbf{m}}$ as functions of magnetization orientation and current direction can be further determined by projecting $\rho_{ij}(\mathbf{m})$ in Eq. (3) to the current direction and transverse direction:

$$\begin{aligned}
\rho_{xx}^{\mathbf{m}}(\theta_{\mathbf{m}}, \phi_{\mathbf{m}}, \phi_{\mathrm{J}}) &= \hat{l}_i(\phi_{\mathrm{J}})\rho_{ij}(\theta_{\mathbf{mc}}, \phi_{\mathbf{mc}})\hat{l}_j(\phi_{\mathrm{J}}) \\
&= A_1 + \frac{1}{2}(A_2 + A_3)\cos 2\phi_{\mathbf{m}} \\
&\quad + \frac{1}{2}(A_2 - A_3)\cos(2\phi_{\mathbf{m}} + 4\phi_{\mathrm{J}}) + A_4\cos(4\phi_{\mathbf{m}} + 4\phi_{\mathrm{J}}) \\
\rho_{xy}^{\mathbf{m}}(\theta_{\mathbf{m}}, \phi_{\mathbf{m}}, \phi_{\mathrm{J}}) &= \hat{t}_i(\phi_{\mathrm{J}})\rho_{ij}(\theta_{\mathbf{mc}}, \phi_{\mathbf{mc}})\hat{l}_j(\phi_{\mathrm{J}}) \\
&= \frac{1}{2}(A_2 + A_3)\sin 2\phi_{\mathbf{m}} - \frac{1}{2}(A_2 - A_3)\sin(2\phi_{\mathbf{m}} + 4\phi_{\mathrm{J}})
\end{aligned} \tag{5}$$

For Fe(001) film, while the current $\mathbf{J}$ is along [100] ($\phi_{\mathrm{J}} = 0°$) or [110] ($\phi_{\mathrm{J}} = 45°$) direction, Eq. (5) can be simplified into

$$\begin{aligned}
\rho_{xx}^{\mathbf{m},[100]}(\theta_{\mathbf{m}}, \phi_{\mathbf{m}}) &= A_1 + A_2\cos 2\phi_{\mathbf{m}} + A_4\cos 4\phi_{\mathbf{m}}, \\
\rho_{xy}^{\mathbf{m},[100]}(\theta_{\mathbf{m}}, \phi_{\mathbf{m}}) &= A_3\sin 2\phi_{\mathbf{m}}, \\
\rho_{xx}^{\mathbf{m},[110]}(\theta_{\mathbf{m}}, \phi_{\mathbf{m}}) &= A_1 + A_3\cos 2\phi_{\mathbf{m}} - A_4\cos 4\phi_{\mathbf{m}}, \\
\rho_{xy}^{\mathbf{m},[110]}(\theta_{\mathbf{m}}, \phi_{\mathbf{m}}) &= A_2\sin 2\phi_{\mathbf{m}}.
\end{aligned} \tag{6}$$

AMR has the combination of a two-fold $\cos 2\phi_{\mathbf{m}}$ term and a four-fold $\cos 4\phi_{\mathbf{m}}$ term, but PHE only has the two-fold $\sin 2\phi_{\mathbf{m}}$ term. The $\cos 2\phi_{\mathbf{m}}$ term in $\rho_{xx}^{\mathbf{m},[100]}$ and the $\sin 2\phi_{\mathbf{m}}$ term in $\rho_{xy}^{\mathbf{m},[110]}$ have the same coefficient of $A_2$, but the $\cos 2\phi_{\mathbf{m}}$ term in $\rho_{xx}^{\mathbf{m},[110]}$ and the $\sin 2\phi_{\mathbf{m}}$ term in $\rho_{xy}^{[100]}$ have the same coefficient of $A_3$. Such reciprocal relation between AMR and PHE with $\mathbf{J}$ along [100] and [110] has been experimentally reported in CoFe films in Ref. [19], and recently emphasized again in Ref. [20] with the magnetization rotating in the film plane. In fact, such reciprocal relation between AMR and PHE with $\mathbf{J}$ along [100] and [110] has already been described by Chen et al. in 1972 [11]. However, in Eq. (6), the coefficients $A_1$, $A_2$, $A_3$, and $A_4$ are $\theta_{\mathbf{m}}$-dependent, and it has never been verified whether the reciprocal relation between AMR and PHE with $\mathbf{J}$ along [100] and [110] are valid for arbitrary $\theta_{\mathbf{m}}$.

In Eq. (6), $\rho_{xx}^{\mathbf{m}}$ for the current along either [100] or [110] contains the four-fold $\cos4\phi_{\mathbf{m}}$ term with the coefficient independent of the current orientation, which has been experimentally confirmed with the magnetization rotating in the film plane [11,17,19]. This phenomenon has been theoretically attributed to the modification of the density of states at the Fermi surface due to the orientation of the magnetization [17]. The magnetization orientation, in turn, introduces an in-plane four-fold symmetry that is consistent with the lattice symmetry. As described in Eqs. (3) and (4), the coefficient of the in-plane four-fold term should be $\theta_{\mathbf{m}}$-dependent. The $\rho_{xy}^{\mathbf{m}}$ term derived from the phenomenological theory only contains the two-fold symmetry, and this result is in contrast with that recently reported in Ref. [20], in which a four-fold term of $\rho_{xy}^{\mathbf{m}}$ was claimed. Such controversy should be further clarified in experiments. Noted that, in most previous studies on AMR and PHE [17,22-28], the field was rotated only in particular planes, such as $xy$, $yz$ or $xz$ planes. In order to better validate the phenomenological theory, it is desired to investigate the angular-dependence on AMR and PHE with arbitrary orientations of $\mathbf{m}$.

We also derived the expressions of $\rho_{xx}^{\mathbf{m}}$ and $\rho_{xy}^{\mathbf{m}}$ considering $O_h$ symmetry for bulk Fe with the current applied in the (001) plane. The derived formula has the same expression as Eq. (6), however, due to the additional symmetry constrains, there are only 6 independent parameters with the constrain of $\rho_1 - 4\rho_2 + \rho_3 - 4\rho_5 + 24\rho_7 = 0$ and $\rho_1 + \rho_3 - 4\rho_7 = 0$ [32]. The number of the free parameters is less than that derived from the theory based on the vector order parameters for galvanomagnetic effects [20], because the in-plane four-fold term for PHE does not exist and the AHE contributions are not considered.

# III. Experiment

To verify the phenomenological theory, we performed the angular-dependent magnetoresistance in a single-crystal Fe(001) films [15,16,33,34]. A 60 nm Fe film was epitaxied on $MgAl_2O_4$(001) substrate in an ultrahigh vacuum magnetron sputtering chamber with a base pressure of $2 \times 10^{-8}$ Torr. The $MgAl_2O_4$ substrate was first cleaned by annealing at 600°C for 2 hours in vacuum. Subsequently, the Fe film was deposited at 300°C with a growth rate of 1.2 nm/min. In order to prevent oxidation, a 4-nm-thick $Al_2O_3$ capping layer was grown on the sample at RT. Due to very low lattice mismatch, the Fe(001) film can be epitaxied $MgAl_2O_4$(001) substrate with very high quality [35]. The XRD spectrum indicates the lattice constant of Fe film is 0.2875 nm, which is slightly higher than that of bulk Fe (0.2866 nm), due to the epitaxial strain [35]. The Laue oscillations around the Fe (002) peak also indicates the sharp interface of Fe film.

The 60 nm Fe film was patterned into many 600 μm × 150 μm Hall bar devices using the standard photolithography and Ar ion bombardment etching processes [17,25-28]. Two kind of Hall bars were prepared with the current direction along Fe[100] and Fe[110]. The transport measurements were conducted in a superconductive magnet system. The sample was mounted on a two-dimensional rotary probe, thus the sample can be rotated to align the field along arbitrary directions relative to the sample. The longitudinal and Hall voltages were measured at 300 K with the standard lock-in technique, and an AC electric current of 1 mA with the frequency of 137.31 Hz was applied during the measurements.

# IV.Results and discussion

## A. Field-dependent measurements

We first measured the field-dependent magnetoresistance by sweeping the field along $x$, $y$, and $z$ directions, as defined in Figure 1(a). Figs. 1(b) and 1(c) show the measured field-dependent $\rho_{xx}$ for $\mathbf{J} \parallel [100]$ and $\mathbf{J} \parallel [110]$ , respectively. For $\mathbf{H} \parallel \hat{\mathbf{z}}$, $\mathbf{m}$ rotates gradually from in-plane direction to out-of-plane direction, and is aligned along the $z$ axis for $H > 2.05$ T. For $\mathbf{H} \parallel \hat{\mathbf{x}}$ (or $\hat{\mathbf{y}}$), the saturation field is less than 0.1 T. After the saturation, the resistivity still changes with the field strength due to the field-suppressed magnon resistance [36-38] or the Lorentz force effect [39-41]. However, the slope of field-dependent MR can be obviously different for the field along different directions. In Fig. 1(b), $\rho_{xx}(H_x)$ is obviously larger than $\rho_{xx}(H_y)$ near zero field, but they become to have the similar value at 10 T, indicating that the measured AMR value can be significantly influenced by the strength of applied field. Both AMR and PHE in Eqs. (1-6) describe the magnetization-orientation-dependent $\rho_{xx}^{\mathbf{m}}(\mathbf{m})$ and $\rho_{xy}^{\mathbf{m}}(\mathbf{m})$, thus the field effect on the measured magnetoresistance should be correctly eliminated.

In general, the total measured resistivity $\rho_{xx}$ in experiments should contain the magnetization-orientation-dependent AMR term $\rho_{xx}^{\mathbf{m}}(\mathbf{m})$ and the field-induced magnetoresistance (FIMR) term $\rho_{xx}^{\mathrm{FIMR}}(\mathbf{H})$, i.e.

$$\rho_{xx}(\mathbf{m}, \mathbf{H}) = \rho_{xx}^{\mathbf{m}}(\mathbf{m}) + \rho_{xx}^{\mathrm{FIMR}}(\mathbf{H}). \tag{7}$$

The FIMR term can be further expressed as [36-41]

$$\rho_{xx}^{\mathrm{FIMR}}(\mathbf{H}) = R_{xx}^{\mathrm{MMR}}(\hat{\mathbf{H}})H + R_{xx}^{\mathrm{OMR}}(\hat{\mathbf{H}})H^2, \tag{8}$$

where the linear term $R_{xx}^{\mathrm{MMR}}(\hat{\mathbf{H}})$ is induced by the suppression of electron-magnon scattering by the magnetic field [36-38], and the quadratic term $R_{xx}^{\mathrm{OMR}}$ is induced by the field-induced Lorentz force [39-41]. Here, $\hat{\mathbf{H}}$ represents the unit vector of magnetic field $\mathbf{H}$ with the strength of $H$. $R_{xx}^{\mathrm{MMR}}$ and $R_{xx}^{\mathrm{OMR}}$ can have different values with different field orientation $\hat{\mathbf{H}}$, but can be fitted using Eqs. (7) and (8) at the saturating condition. The fitted curves are shown as red lines in Figs. 1(b) and 1(c). The fitted coefficient $R_{xx}^{\mathrm{OMR}}(\hat{\mathbf{x}})$ is significantly smaller than $R_{xx}^{\mathrm{OMR}}(\hat{\mathbf{y}})$ and $R_{xx}^{\mathrm{OMR}}(\hat{\mathbf{z}})$, because the Lorentz force vanishes for the field along the current.

$\rho_{xx}^{\mathbf{m}}$ represents the longitudinal resistivity without the field effect, which can be obtained through the fitting with Eq. (7). For both current directions, the Fe films have the relation $\rho_{xx}^{\mathbf{m}}(\hat{\mathbf{x}}) > \rho_{xx}^{\mathbf{m}}(\hat{\mathbf{z}}) > \rho_{xx}^{\mathbf{m}}(\hat{\mathbf{y}})$. The relation $\rho_{xx}^{\mathbf{m}}(\hat{\mathbf{x}}) > \rho_{xx}^{\mathbf{m}}(\hat{\mathbf{y}})$ is consistent with the conventional AMR theory in thin films [14,15,39]. For bulk Fe, while the current is along the $\langle 100 \rangle$ direction, the resistance should be the same for the magnetization along y or z axes. In our measurements, the resistivity components $\rho_{xx}^{\mathbf{m}}(\hat{\mathbf{y}})$ and $\rho_{xx}^{\mathbf{m}}(\hat{\mathbf{z}})$ exhibit different values, reflecting the lack of equivalence between the y and z axes. The underlying microscopic mechanism of $\rho_{xx}^{\mathbf{m}}(\hat{\mathbf{y}}) \neq \rho_{xx}^{\mathbf{m}}(\hat{\mathbf{z}})$ can be attributed to geometry size effects [42-44] or epitaxial strain [45-47].

Figs. 1(d) and 1(e) show the measured Hall signals as a function of field strength with three typical field orientations for $\mathbf{J} \parallel [100]$ and $\mathbf{J} \parallel [110]$. For $\mathbf{H} \parallel \hat{\mathbf{z}}$, only AHE signal can be measured, which is antisymmetric to the field. From the AHE curve, we can obtain the saturation field $\mu_0 H_s$, which is 2.05 T in our measurement. For the field applied in the film plane, the AHE contribution on the Hall signal should be zero, and there is PHE signal symmetric to the field. While the field is applied along arbitrary orientation, the total Hall signal consists of both AHE and PHE signals, which can be separated through the symmetry analysis. In this study, we only focus on the PHE signal symmetric to the field. Since the PHE signal originates from AMR [9-11], which contains both $\mathbf{m}$-induced and $\mathbf{H}$-induced terms in Eq. (7), the measured PHE signal $\rho_{xy}^{\mathrm{PHE}}$ also consists of the $\mathbf{m}$-induced contribution and $\mathbf{H}$-induced contribution, i.e.

$$\rho_{xy}^{\mathrm{PHE}}(\mathbf{m}, \mathbf{H}) = \rho_{xy}^{\mathbf{m}}(\mathbf{m}) + \rho_{xy}^{\mathrm{FIMR}}(\mathbf{H}). \quad (9)$$

$\rho_{xy}^{\mathrm{FIMR}}(\mathbf{H})$ is the PHE signal induced by anisotropic FIMR, thus also consists of the linear term $R_{xy}^{\mathrm{MMR}}(\hat{\mathbf{H}})$ induced by the suppression of electron-magnon scattering by the magnetic field [36-38], and the quadratic term $R_{xy}^{\mathrm{OMR}}$ induced by the field-induced Lorentz force [39-41], similar to Eq. (8).

## B. Angular-dependent measurements

We also performed the measurement of $\rho_{xx}$ and $\rho_{xy}$ with the field rotating in the $xy$, $yz$, and $xz$ planes with a constant field strength. Fig. 1(a) defines the orientation angles of $\mathbf{H}$ labeled as $\alpha_{\mathbf{H}}$, $\beta_{\mathbf{H}}$, and $\gamma_{\mathbf{H}}$ in each rotation plane. Fig. 2(a) shows the field-angle dependence of $\rho_{xx}$ for the device with $\mathbf{J} \parallel [100]$ with a 4 T rotating field. The hysteresis loop measurements show that the magnetization can be fully saturated by a 0.05 T in-plane field. For the in-plane field-rotating measurement, it is expected that the magnetization is always aligned with the 4 T field, thus the $\alpha_{\mathbf{H}}$-dependent data can be well fitted by Eq. (6) if assuming $\theta_{\mathbf{m}} = 90°$ and $\phi_{\mathbf{m}} = \alpha_{\mathbf{H}}$. However, Figs. 1(d) and 1(e) show that the out-of-plane saturating field is 2.05 T, thus there is a misalignment between $\mathbf{m}$ and $\mathbf{H}$ for the field rotating in the $yz$ and $xz$ planes, then the $\beta_{\mathbf{H}}$- and $\gamma_{\mathbf{H}}$-dependent $\rho_{xx}$ in Fig. 2(a) cannot be well fitted by Eq. (6). The polar angle $\theta_{\mathbf{m}}$ of $\mathbf{m}$ can be calculated through the field angle $\theta_{\mathbf{H}}$ . By minimizing total energy including the uniaxial magnetic anisotropy energy and Zeeman energy, the relation between $\theta_{\mathbf{M}}$ and $\theta_{\mathbf{H}}$ can be determined by

$$\sin(\theta_{\mathbf{m}} - \theta_{\mathbf{H}}) = \frac{H_k}{2H}\sin 2\theta_{\mathbf{m}}. \tag{10}$$

According to Fig. 1, the out-of-plane anisotropy field $H_{\mathrm{k}}$ can be estimated as 2.05 T, thus the polar angle $\theta_{\mathbf{m}}$ as a function of $\theta_{\mathbf{H}}$ with a fixed field can be calculated through Eq. (10). Fig. 2(b) replots the $\beta_{\mathbf{m}}$- and $\gamma_{\mathbf{m}}$-dependent $\rho_{xx}$, which can be well fitted by Eq. (6).

The field-orientation dependent Hall signals with the field rotating in the $xy$, $yz$, and $xz$ planes were also measured. Figs. 2(c) and 2(d) depict the Hall data as a function of $\theta_{\mathbf{H}}$ and $\theta_{\mathbf{m}}$, respectively. For the field rotating in $xy$ plane, only PHE signal can be measured, and a $\sin 2\phi_{\mathbf{m}}$ term can well fit the data. For the field rotating in $yz$ and $xz$ planes, the PHE signal should be zero, and only AHE signal can be measured. The AHE signal cannot be fitted with the $\cos\theta_{\mathbf{H}}$ function in Fig. 2(c), but can be well fitted with the $\cos\theta_{\mathbf{m}}$ function in Fig. 2(d).

We also performed the similar measurements and analysis on the device with $\mathbf{J} \parallel [110]$. Figs. 2(e) and 2(f) show the measured magnetization-angle dependence of $\rho_{xx}$ and $\rho_{xy}$. Both $\rho_{xx}$ and $\rho_{xy}$ curves can be well fitted using Eq. (6) derived from the phenomenological theory.

Fig. 1 has demonstrated that the field-induced changing of $\rho_{xx}$ significantly depends on the field orientation, which can make the measured AMR strongly depending on the field strength. AMR typically refers to changes in magnetoresistance induced by variations in the magnetization orientation. To comprehensively understand the AMR properties, especially when the magnetic field is rotating within the $yz$ and $xz$ planes, it is essential to apply a strong field exceeding the saturation field [17,19,20,25-28]. Therefore, to fully comprehend the angular dependence of AMR, it is necessary to eliminate the anisotropic FIMR contribution, thereby isolating the intrinsic relationship of $\rho_{xx}^{\mathbf{m}}(\mathbf{m})$.

Fig. 3(a) compares the field-angle dependent $\rho_{xx}(\alpha_{\mathbf{m}})$ with the applied fields of 4 T and 6 T from the device with $\mathbf{J} \parallel [100]$. A significant difference between two curves can be observed, which is induced by the anisotropic FIMR [16,39-41]. From the field-dependent measurements in Fig. 1, the linear ($R_{xx}^{\mathrm{OMR}}$) and quadratic ($R_{xx}^{\mathrm{MMR}}$) FIMR coefficients with the field along $\hat{\mathbf{x}}$, $\hat{\mathbf{y}}$ and $\hat{\mathbf{z}}$ axes can be obtained. For the arbitrary field orientation $\hat{\mathbf{H}}$, $R_{xx}^{\mathrm{FIMR}}(\hat{\mathbf{H}})$ can be calculated by the field components along three principal axes. Since the FIMR coefficients should be symmetric to the field, in simplest case, the FIMR coefficient can be described as $R_{xx}^{\mathrm{FIMR}}(\hat{\mathbf{H}}) = R_{xx}^{\mathrm{FIMR}}(\hat{\mathbf{x}})\hat{\mathbf{H}}_x^2 + R_{xx}^{\mathrm{FIMR}}(\hat{\mathbf{y}})\hat{\mathbf{H}}_y^2 + R_{xx}^{\mathrm{FIMR}}(\hat{\mathbf{z}})\hat{\mathbf{H}}_z^2$, with $\hat{\mathbf{H}}_x$, $\hat{\mathbf{H}}_y$ and $\hat{\mathbf{H}}_z$ are the projection components of $\hat{\mathbf{H}}$ on the three principle axes. Thus, if describing the field orientation with $\theta_{\mathbf{H}}$ and $\phi_{\mathbf{H}}$, the FIMR coefficients with arbitrary field orientation can be expressed as:

$$\begin{aligned} R_{xx}^{\mathrm{OMR}}(\hat{\mathbf{H}}) &= R_{xx}^{\mathrm{OMR}}(\hat{\mathbf{x}})\sin^2\theta_{\mathbf{H}}\cos^2\phi_{\mathbf{H}} + R_{xx}^{\mathrm{OMR}}(\hat{\mathbf{y}})\sin^2\theta_{\mathbf{H}}\sin^2\phi_{\mathbf{H}} + R_{xx}^{\mathrm{OMR}}(\hat{\mathbf{z}})\cos^2\theta_{\mathbf{H}}, \\ R_{xx}^{\mathrm{MMR}}(\hat{\mathbf{H}}) &= R_{xx}^{\mathrm{MMR}}(\hat{\mathbf{x}})\sin^2\theta_{\mathbf{H}}\cos^2\phi_{\mathbf{H}} + R_{xx}^{\mathrm{MMR}}(\hat{\mathbf{y}})\sin^2\theta_{\mathbf{H}}\sin^2\phi_{\mathbf{H}} + R_{xx}^{\mathrm{MMR}}(\hat{\mathbf{z}})\cos^2\theta_{\mathbf{H}}. \end{aligned} \quad (11)$$

Thus, $\rho_{xx}^{\text{FIMR}}(\mathbf{H})$ can be fully determined with Eqs. (8) and (11). After removing the FIMR contributions, we can obtain the intrinsic $\rho_{xx}^{\mathbf{m}}(\mathbf{m})$ signals. As shown in Fig. 3(b), the determined $\rho_{xx}^{\mathbf{m}}(\mathbf{m})$ curves with the rotating field of 6 T and 4 T are completely same, indicating that the FIMR contribution has been successfully eliminated.

Since PHE originates from AMR [9-11], the AMR induced by strong magnetic field can induce the change of PHE signal, which also includes the linear term and the quadratic term. So, the field induced change of $\rho_{xy}$ can be described as $\rho_{xy}^{\text{FIMR}}(\mathbf{H}) = R_{xy}^{\text{MMR}}(\widehat{\mathbf{H}})H + R_{xy}^{\text{OMR}}(\widehat{\mathbf{H}})H^2$, similar to that of $\rho_{xx}$ in Eq. (8). In general, the maximum PHE signal is observed with the in-plane field aligned along the direction $45°$ away from the current, which is symmetric to the magnetic field [9-11], so we assume both $R_{xy}^{\text{OMR}}(\widehat{\mathbf{H}})$ and $R_{xy}^{\text{MMR}}(\widehat{\mathbf{H}})$ have the following simple angular-dependence :

$$\begin{aligned} R_{xy}^{\text{OMR}}(\widehat{\mathbf{H}}) &= R_{xy,0}^{\text{OMR}}\widehat{\mathbf{H}}_{\parallel}^{2}\sin 2\phi_{\mathbf{H}} = R_{xy,0}^{\text{OMR}}\sin^2\theta_{\mathbf{H}}\sin 2\phi_{\mathbf{H}}, \\ R_{xy}^{\text{MMR}}(\widehat{\mathbf{H}}) &= R_{xy,0}^{\text{MMR}}\widehat{\mathbf{H}}_{\parallel}^{2}\sin 2\phi_{\mathbf{H}} = R_{xy,0}^{\text{MMR}}\sin^2\theta_{\mathbf{H}}\sin 2\phi_{\mathbf{H}}. \end{aligned} \tag{12}$$

$R_{xy,0}^{\text{OMR}}$ and $R_{xy,0}^{\text{MMR}}$ can be determined from the field-dependent measurements of $\rho_{xy}$ for the in-plane field sweeping with 45° away from the current, i.e. $(\theta_{\mathbf{H}}, \phi_{\mathbf{H}}) = (90°, 45°)$ [32].

Fig. 3(c) shows the typical $\rho_{xy}^{\text{PHE}}$ curves for the device with $\mathbf{J} \parallel [110]$ with the field rotating in the $xy$ plane. The two curves, measured at the field strengths of 6 T and 4 T, exhibit significant difference. Fig. 3(d) presents the two nearly identical curves after subtracting the $\rho_{xy}^{\text{FIMR}}(\mathbf{H})$ contributions, which further validate the methodology for determining the field-induced contributions on AMR and PHE, as outlined by Eqs. (11) and (12). Thus, the intrinsic $\rho_{xx}^{\mathbf{m}}(\mathbf{m})$ and $\rho_{xy}^{\mathbf{m}}(\mathbf{m})$, which are induced by changes in the magnetization orientation, can be successfully extracted by subtracting the field-induced signal from the measured data $\rho_{xx}(\mathbf{m}, \mathbf{H})$ and $\rho_{xy}^{\text{PHE}}(\mathbf{m}, \mathbf{H})$. This approach is crucial for further validating the phenomenological theory of galvanomagnetic effects.

## C. Full angular-mapping of longitudinal resistivity and planar Hall resistivity

Next, we measured the $\rho_{xx}^{\mathbf{m}}(\mathbf{m})$ and $\rho_{xy}^{\mathbf{m}}(\mathbf{m})$ with arbitrary magnetization orientations $(\theta_{\mathbf{m}}, \phi_{\mathbf{m}})$ and compared the experimental results with the phenomenological theory. Fig. 4(a) shows the schematics for measuring AMR and PHE with the field applied across the entire angular space. The field was first tilted away from the normal direction with the polar angle $\theta_{\mathbf{H}}$, then the sample was rotated azimuthally by varying $\phi_{\mathbf{H}}$. Since the applied field was usually larger than 3 T, the azimuthal angle $\phi_{\mathbf{m}}$ is expected to be same as $\phi_{\mathbf{H}}$. However, due to strong shape anisotropy, $\theta_{\mathbf{m}}$ could be significantly different from $\theta_{\mathbf{H}}$, and their relation can be calculated using Eq. (10). In this measurement, the sweeping steps for both $\phi_{\mathbf{H}}$ and $\theta_{\mathbf{H}}$ are 5º. The rotation range of $\phi_{\mathbf{H}}$ is only 320º limited by the rotary probe. Our measurements were conducted with $\theta_{\mathbf{H}}$ only between $0°$ and $90°$, since Eq. (4) shows that the results for $\theta_{\mathbf{H}}$ between $180°$ and $90°$ should be the same.

Figures 4(b) and 4(d) show the measured $\phi_{\mathbf{m}}$-dependent longitudinal resistivities with different $\theta_{\mathbf{H}}$ from the devices with $\mathbf{J} \parallel [100]$ and $\mathbf{J} \parallel [110]$, respectively. The rotating field strength is 6 T. The field-induced magnetoresistance $\rho_{xx}^{\mathrm{FIMR}}(\mathbf{H})$ has been subtracted utilizing equations (8) and (11). In Fig. 4, we use $\Delta\rho_{xx}^{\mathbf{m}}(\mathbf{m}) = \rho_{xx}^{\mathbf{m}}(\mathbf{m}) - \rho_{xx}^{\mathbf{m}}(\hat{\mathbf{z}})$ to represent the change of the longitudinal resistivity, and $\rho_{xx}^{\mathbf{m}}(\hat{\mathbf{z}})$ is the resistivity with the field along the *z* axis. According to Eq. (6), for a fixed $\theta_{\mathbf{m}}$, all the $\rho_{xx}^{\mathbf{m}}(\phi_{\mathbf{m}})$ curves can be well fitted with a two-fold term and a four-fold term. Figs. 4(c) and 4(e) present the $\phi_{\mathbf{m}}$-dependent planar Hall resistivities $\rho_{xy}^{\mathrm{PHE}}(\mathbf{m})$ measured in two devices, which only contain a two-fold term. The field-induced contribution $\rho_{xy}^{\mathrm{FIMR}}(\mathbf{H})$ was subtracted utilizing Eq. (12).

The experimental results in Fig. 4 are well consistent with the phenomenological theory, indicated by the red fitting curves. All the $\phi_{\mathbf{m}}$-dependent $\rho_{xx}^{\mathbf{m}}$ and $\rho_{xy}^{\mathbf{m}}$ curves can be well fitted by Eq. (6). $\Delta\rho_{xx}^{\mathbf{m}}$ with $\mathbf{J} \parallel [100]$ and $\rho_{xy}^{\mathbf{m}}$ with $\mathbf{J} \parallel [110]$ show the similar amplitude, and the amplitudes of $\rho_{xy}^{\mathbf{m}}$ with $\mathbf{J} \parallel [100]$ and $\Delta\rho_{xx}^{\mathbf{m}}$ with $\mathbf{J} \parallel$ [110] show the similar value, which agree well with the reciprocal relation shown in Eq. (6). Though the fitting, we can obtain the $\theta_{\mathbf{H}}$-dependent parameters $A_1$, $A_2$, $A_3$, and $A_4$. Fig. 4(f) shows the $\theta_{\mathbf{m}}$-dependence of $A_2$ and $A_3$ fitted from both $\Delta\rho_{xx}^{\mathbf{m}}$ and $\rho_{xy}^{\mathbf{m}}$ curves in both devices, and the relation between $\theta_{\mathbf{H}}$ and $\theta_{\mathbf{m}}$ is determined using Eq. (10). The reciprocal relation between AMR and PHE in two devices are valid for all magnetization angles $\theta_{\mathbf{m}}$. The dependence of $A_2$ and $A_3$ on $\theta_{\mathbf{m}}$, as illustrated in Fig. 4(f) can be effectively described by Eq. (4) and is well fitted to the data. The excellent fit observed for both $\theta_{\mathbf{m}}$-dependence and $\phi_{\mathbf{m}}$-dependence of $\Delta\rho_{xx}^{\mathbf{m}}$ and $\rho_{xy}^{\mathbf{m}}$ in the Fe(001) film demonstrates that the magnetization-angle-dependent galvanomagnetic effects are accurately captured by the phenomenological theory.

To further substantiate that the phenomenological theory accurately describes the magnetization-dependence of galvanomagnetic effects, we have depicted the $\theta_{\mathbf{m}}$-dependence and $\phi_{\mathbf{m}}$-dependence of $\Delta\rho_{xx}^{\mathbf{m}}$ and $\rho_{xy}^{\mathbf{m}}$ in 3-dimensional plot, as illustrated by the red data points in Fig. 5. The dataset presented here can be effectively fitted using a single set of coefficients, which are detailed in Table 1. The curved surface plots are the fitted angular dependence of $\Delta\rho_{xx}^{\mathbf{m}}$ and $\rho_{xy}^{\mathbf{m}}$ using Eq. (6). The shaded area at the bottom in each figure indicates the discrepancy between the experimental data and the fitting results. The low fitting residue indicates that phenomenological theory is enough to describe AMR and PHE with arbitrary orientation of magnetization in the Fe(001) film. The similar angular-dependent measurement was also performed with a rotating field of 4 T, and the experimental data can also be well fitted by one single set of coefficients shown in Table 1. The fitted results with the fields of 4 T and 6 T are very similar in comparison with the fitting errors, which indicates that the field effects on AMR and PHE have been successfully eliminated. Note that only 8 parameters $\rho_0, \rho_1, \ldots, \rho_7$ can be quantified in our experiments, however there are 13 independent parameters $a_{ij}, a_{klij}, a_{klmnij}$ in Eq. (2) based on the symmetry analysis, which cannot be fully quantified.

The full-angular mappings of $\rho_{xy}^{\mathbf{m}}$ depicted in Figs. 5(b) and 5(d) clearly demonstrate the absence of a four-fold term in the PHE signal, which is in full accord with Eq. (6) derived by the phenomenological theory for systems exhibiting cubic or $C_{4v}$ symmetry. However, this observation contradicts the recent theoretical description based on the vector order parameters [20]. Our in-depth analysis suggests that within a system possessing $C_4$ symmetry, the presence of a four-fold term in $\rho_{xy}^{\mathbf{m}}$ can be rigidly prohibited by the Onsager relation and the operation with $90°$ rotation around the $z$ axis.

Figure 6 elucidates the symmetry operations with the Onsager relation confirming the absence of the four-fold term in the planar Hall resistivity $\rho_{xy}^{\mathbf{m}}$. Fig. 6(a) depicts the general configuration for measuring the PHE, with the electric current $\mathbf{J}$ directed along the $x$ axis, the transverse electric field $\mathbf{E}_H$ along the $y$ axis, and the in-plane magnetization $\mathbf{m}$ at an angle $\phi_m$ from $\mathbf{J}$ within the film. According to the Onsager reciprocity relation $\rho_{xy}(\mathbf{m}) = \rho_{yx}(-\mathbf{m})$, inverting the magnetization and swapping the directions of $\mathbf{J}$ and $\mathbf{E}_H$ should yield an identical $\rho_{xy}^{\mathbf{m}}$ signal, as shown in Fig. 6(b). In a $C_4$ symmetric system, a 90° counterclockwise rotation about the $z$ axis should result in an equivalent $\rho_{xy}^{\mathbf{m}}$ value, as illustrated in Fig. 6(c). In both Figs. 6(a) and 6(c), $\mathbf{J}$ remains same while $\mathbf{E}_H$ reverses, and $\mathbf{m}$ exhibits a 90° angular difference, establishing the relation $\rho_{xy}^{\mathbf{m}}(\phi_{\mathbf{m}}) = -\rho_{xy}^{\mathbf{m}}\left(\phi_{\mathbf{m}} + \frac{\pi}{2}\right)$. This relationship effectively precludes the presence of the $\sin 4\phi_{\mathbf{m}}$ term in $\rho_{xy}^{\mathbf{m}}$. Therefore, the occurrence of a four-fold term in the PHE in a FM film with $C_{4v}$ symmetry is deemed highly unlikely.

We also noticed that the presence of a $\sin 4\phi_{\mathbf{m}}$ term in $\rho_{xy}^{\mathbf{m}}$ could be conceivable if the four-fold rotation symmetry is broken. The minor four-fold PHE term identified in Ref. [20] through fitting experimental data might be a consequence of this symmetry breaking. Such a disruption in symmetry could potentially stem from surface miscut [48], given that the substrate may not perfectly align with the (001) plane. Additionally, atomic defects, which are common in single-crystal samples, can also affect the symmetry. While these defects locally break symmetry, the long-range four-fold symmetry of the crystal is preserved, since the Hall bar devices, with dimensions on the order of hundreds of microns, average out the effects of such defects. Therefore, atomic defects are unlikely to account for the nearly zero value of the four-fold term in the PHE curve. However, the net contribution of atomic defects may introduce a two-fold dependence of AMR on $\phi_{\mathbf{m}}$, independent of current orientation. Nevertheless, the band structure effect is expected to dominate the contribution to AMR in single-crystal systems. Even in disordered FeCo alloys [19], the AMR anisotropy ratio, which can reach up to 4200%, closely aligns with first-principles calculations based on the band structure.

Our analysis indicates that if the $C_4$ symmetry is reduced to the $C_2$ symmetry, an additional term, $\rho_8 \sin^4\theta_{\mathbf{m}} \sin 4\phi_{\mathbf{m}}$, would emerge in the expression for $\rho_{xy}^{\mathbf{m}}(\mathbf{m})$ in Eq. (6). Incorporating this additional four-fold term into the fitting for the data $\rho_{xy}^{\mathbf{m}}(\mathbf{m})$ depicted in Fig. 5, the coefficient $\rho_8$ is fitted to be $-0.008 \pm 0.084\ \mathrm{n\Omega \cdot cm}$. This fitted value falls below the margin of error, thereby substantiating the absence of a four-fold PHE term in the Fe(001) films.

Both the conventional phenomenological theory established by Döring [12] and the recent theory of Miao et al. [20] can provide good descriptions on the angular-dependence of the anisotropic galvanomagnetic effects in single crystal systems. In both theories, the angular dependence of galvanomagnetic effects is analyzed based on the symmetry analysis, and the primary distinction between the two theories is the approach to modeling the rank-2 resistivity tensor $\overleftrightarrow{\rho}$. In the conventional phenomenological theory, the matrix elements $\rho_{ij}$ are expressed as a MacLaurin's series expansion, as illustrated in Eq. 1. In contrast, Miao et al. decompose $\overleftrightarrow{\rho}$ into a combination of a scalar term, seven vectors and ten traceless symmetric tensors of ranks 2. The coefficients of these vectors and tensors are magnetization-direction-dependent and can also be expanded as series in terms of $\mathbf{m}$. By conducting a proper symmetry analysis based on crystal structures, both theories should yield the same angular dependence for anisotropic galvanomagnetic effects. We can demonstrate this equivalence in systems with $O_h$ symmetry, with the additional consideration of the Onsager relation [32]. In magnetic thin films, the bulk $O_h$ symmetry typically reduces to $C_{4v}$ symmetry. According to conventional phenomenological theory, the matrix elements in crystalline systems with varying symmetries have been extensively analyzed [17-19,31]. This analysis should yield universal angular dependencies for both longitudinal and transverse resistivity in most crystal systems with different symmetries.

Our experimental findings on Fe films substantiate that the conventional phenomenological theory adequately describes the angular dependence of anisotropic galvanomagnetic effects, with all coefficients quantitatively determined. However, such analytical models based on symmetry analysis do not elucidate the physical origins of these coefficients in real material systems. It is reasonable to assume that AMR in crystalline magnetic materials is influenced by their electronic band structure. Consequently, to unravel the microscopic mechanisms underlying AMR, it is imperative to employ first-principles calculations that simulate transport properties across varying current directions and magnetization orientations [19,21,49-52]. In CoFe(001) alloy films [19], theoretical calculations have revealed that the topological nodal line within the electronic band structures significantly contributes to the large current-orientation-dependent AMR. Similarly, theoretical studies in FePt(001) films [21] have linked the observed in-plane four-fold term to the change of the density of states near the Fermi surface under rotation of the magnetization. Our experimental results highlight the need for further theoretical explorations that consider the band structure of Fe to elucidate the fundamental quantum origins of each coefficient related to anisotropic galvanomagnetic effects. Such insights are crucial for a comprehensive understanding of the intrinsic spin-dependent transport characteristics in magnetic materials.

## V. Conclusion

In conclusion, our study comprehensively demonstrates that the full angular dependence of galvanomagnetic effects in single-crystal Fe(001) films is accurately captured by the conventional phenomenological theory based on symmetry analysis. Through meticulous analysis, we separated the intrinsic galvanomagnetic effects, $\rho_{xx}^{\mathbf{m}}$ and $\rho_{xy}^{\mathbf{m}}$, due to magnetization orientation changes from those induced by the applied magnetic field. Our comprehensive experimental analysis across the full angular space confirms that only eight independent and intrinsic parameters are sufficient to describe the angular dependencies of $\rho_{xx}^{\mathbf{m}}$ and $\rho_{xy}^{\mathbf{m}}$ in a ferromagnetic film with $C_{4v}$ symmetry. The excellent agreement between our experimental data and the phenomenological theory, parameterized by these eight coefficients, validates the theory's ability to describe both AMR and PHE in Fe(001) films for arbitrary magnetization orientations. Additionally, our findings underscore the absence of the four-fold term in PHE for systems with $C_{4v}$ symmetry, a result consistent with the constraints imposed by the Onsager reciprocal relation and the four-fold rotation symmetry. While our study focuses on Fe(001) films, the phenomenological theory is likely applicable to other crystal systems with varying symmetries. However, further research is needed to explore the microscopic mechanisms underlying the angular-dependent coefficients derived from the phenomenological theory, particularly the connection between the band structure and the intrinsic galvanomagnetic effects. This work lays the foundation for a more profound understanding of the fundamental physics of galvanomagnetic effects in high-quality ferromagnetic materials.

## ACKNOWLEDGEMENTS

The work was supported by the National Key Research and Development Program of China (Grant No. 2022YFA1403300), the National Natural Science Foundation of

China (Grant No. 12274083, No. 12434003, No. 12221004 and No. 12174028), the Shanghai Municipal Science and Technology Major Project (Grant No. 2019SHZDZX01), and the Shanghai Municipal Science and Technology Basic Research Project (Grant No. 22JC1400200 and No. 23dz2260100).

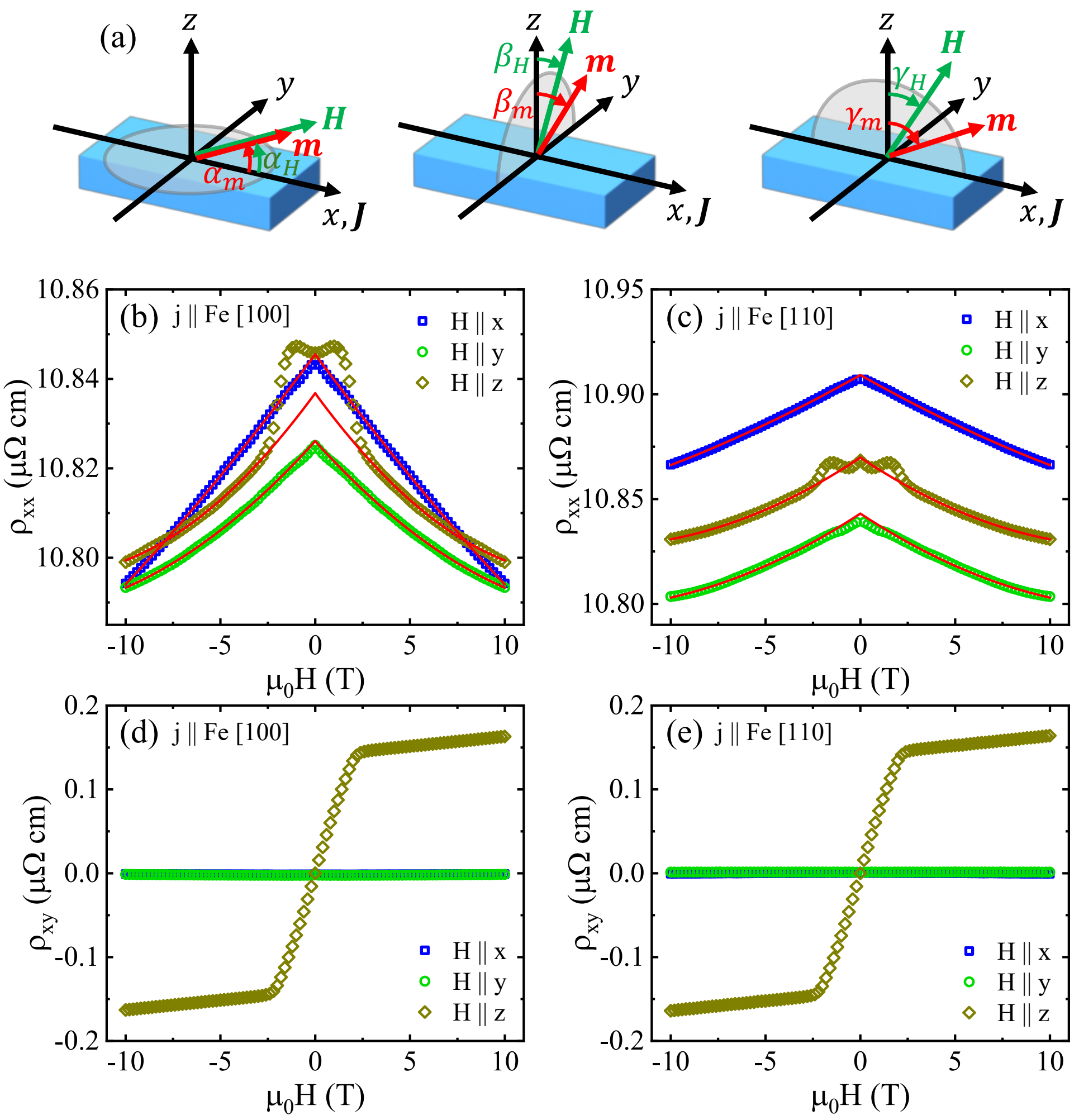

**Fig. 1.** (a) Schematic of the measurement geometry and angle definitions. The current direction is defined along the x-axis, with rotation angles in the xy, yz, and xz planes denoted as α, β, and γ, respectively. (b-c) Field dependences of longitudinal resistivity $\rho_{xx}$ for the devices with (b) $\boldsymbol{J} \parallel [100]$ and (c) $\boldsymbol{J} \parallel [110]$, with the field applied along three primary axes. The lines in (b-c) are the fitting curves with Eq. (8). (d-e) Field dependences of Hall resistivity $\rho_{xy}$ for the devices with (d) $\boldsymbol{J} \parallel [100]$ and (e) $\boldsymbol{J} \parallel [110]$.

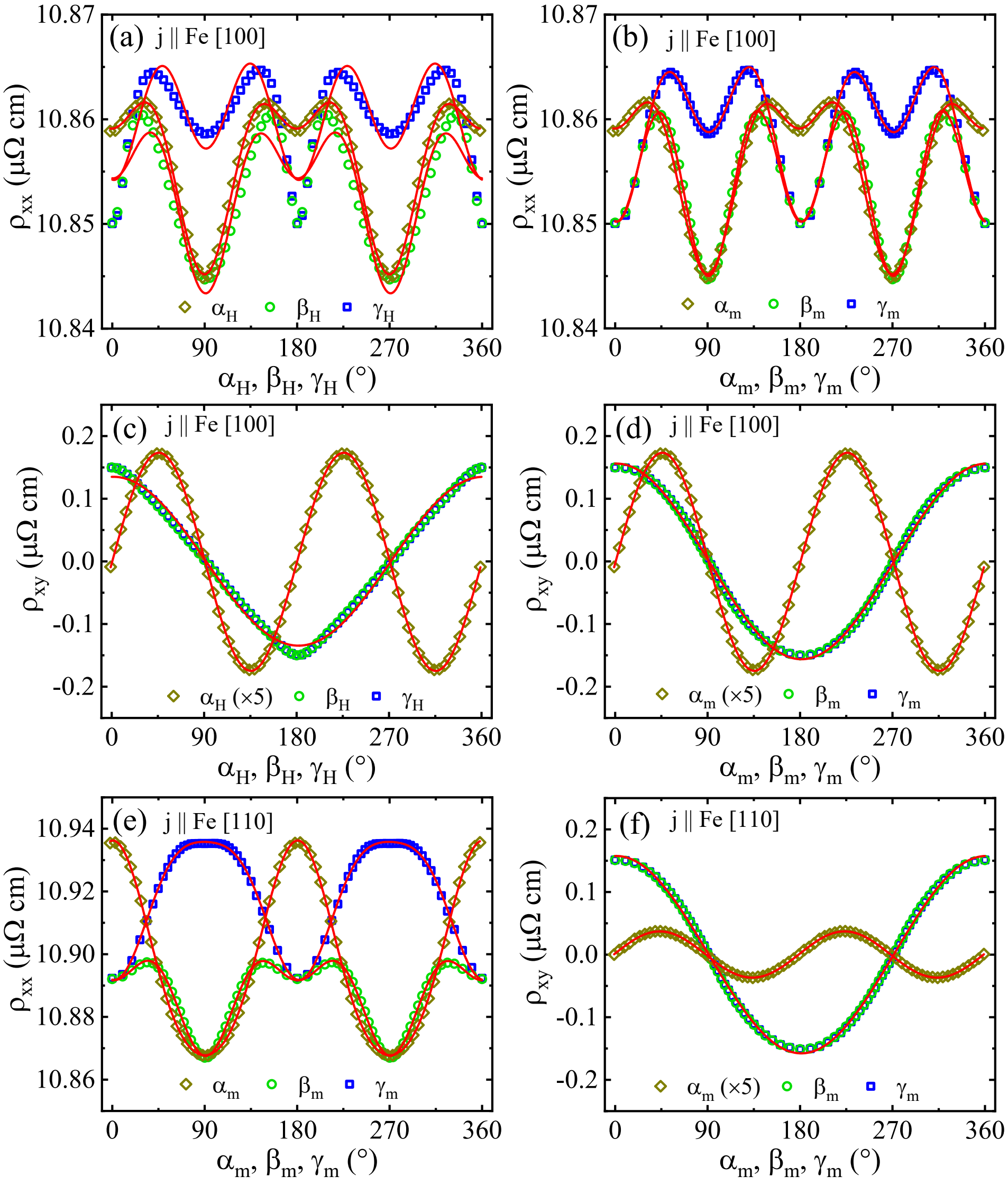

**Fig. 2.** (a-b) Angular dependence of $\rho_{xx}$ in the device with $\boldsymbol{J} \parallel [100]$, with respect to the orientation angles of (a) the magnetic field and (b) magnetization, respectively. (c-d) Angular dependence of $\rho_{xy}$ in the device with $\boldsymbol{J} \parallel [100]$ with respect to the orientation angles of (c) the magnetic field and (d) magnetization, respectively. (e-f) Angular dependence of (e) $\rho_{xx}$ and (f) $\rho_{xy}$ in the $\boldsymbol{J} \parallel [110]$ device with respect to orientation angles of the magnetization, respectively. For clarity, the $\alpha_{\boldsymbol{H}}$- and $\alpha_{\boldsymbol{m}}$-dependent $\rho_{xy}$ curves of in (c-d) and (f) are multiplied by 5. All the solid lines are the fitting curves with Eq. (6).

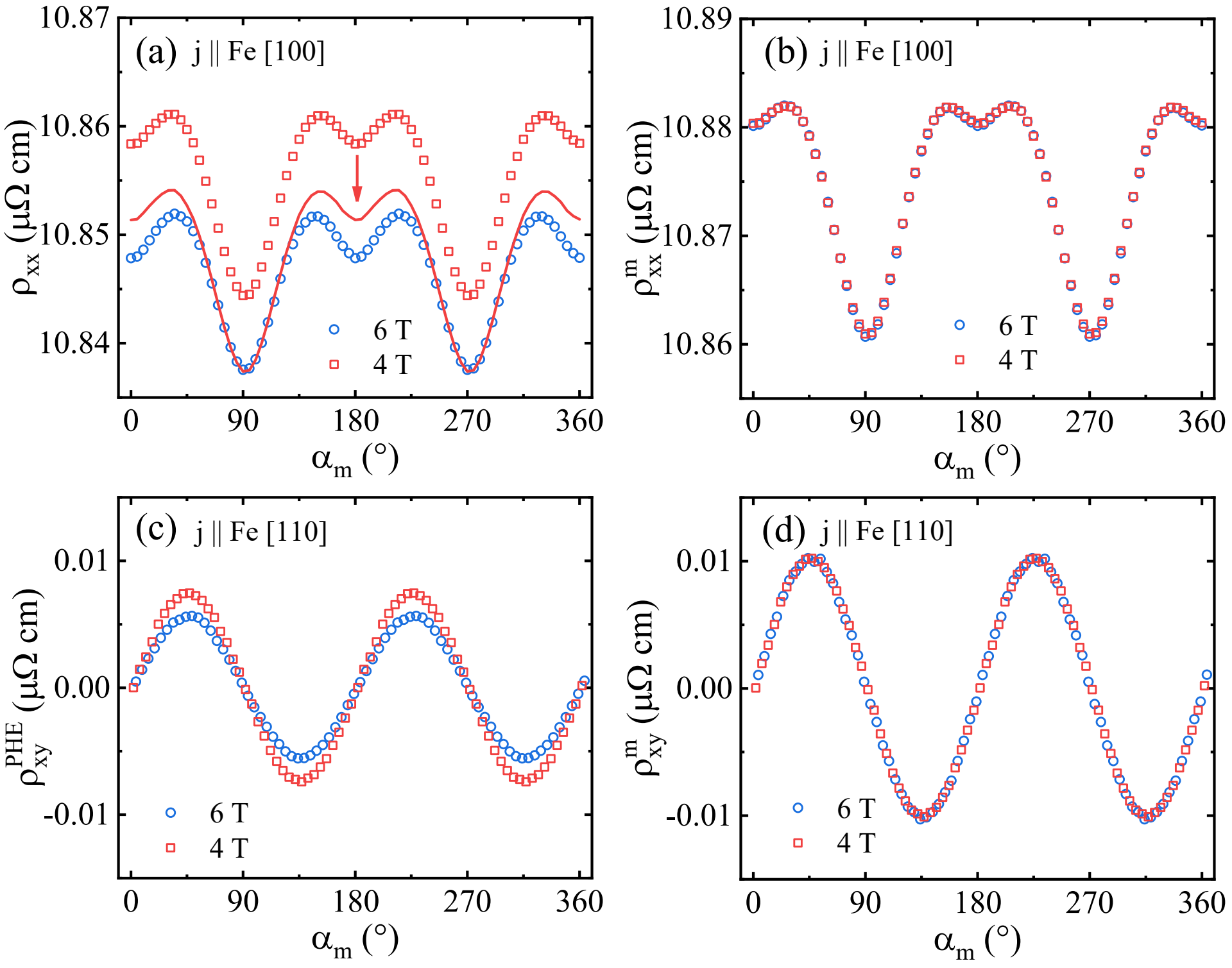

**Fig. 3.** (a) The $\alpha_{\boldsymbol{m}}$-dependence of $\rho_{xx}$ measured with field strengths of 4 T and 6 T. The red line is the shift $\rho_{xx}$ curve of 4 T to better emphasize the anisotropy of the field-induced signals $\rho_{xx}^{FIMR}$. (b) The intrinsic magnetization-induced $\rho_{xx}^{\boldsymbol{m}}$ with $\rho_{xx}^{FIMR}$ subtracted to isolate the magnetization effect. (c) The $\alpha_{\boldsymbol{m}}$-dependence of $\rho_{xy}^{PHE}$ measured with different field strengths. (d) The intrinsic magnetization-induced $\rho_{xx}^{\boldsymbol{m}}$ signals, with the field-induced signals $\rho_{xy}^{FIMR}$ subtracted to isolate the magnetization effects.

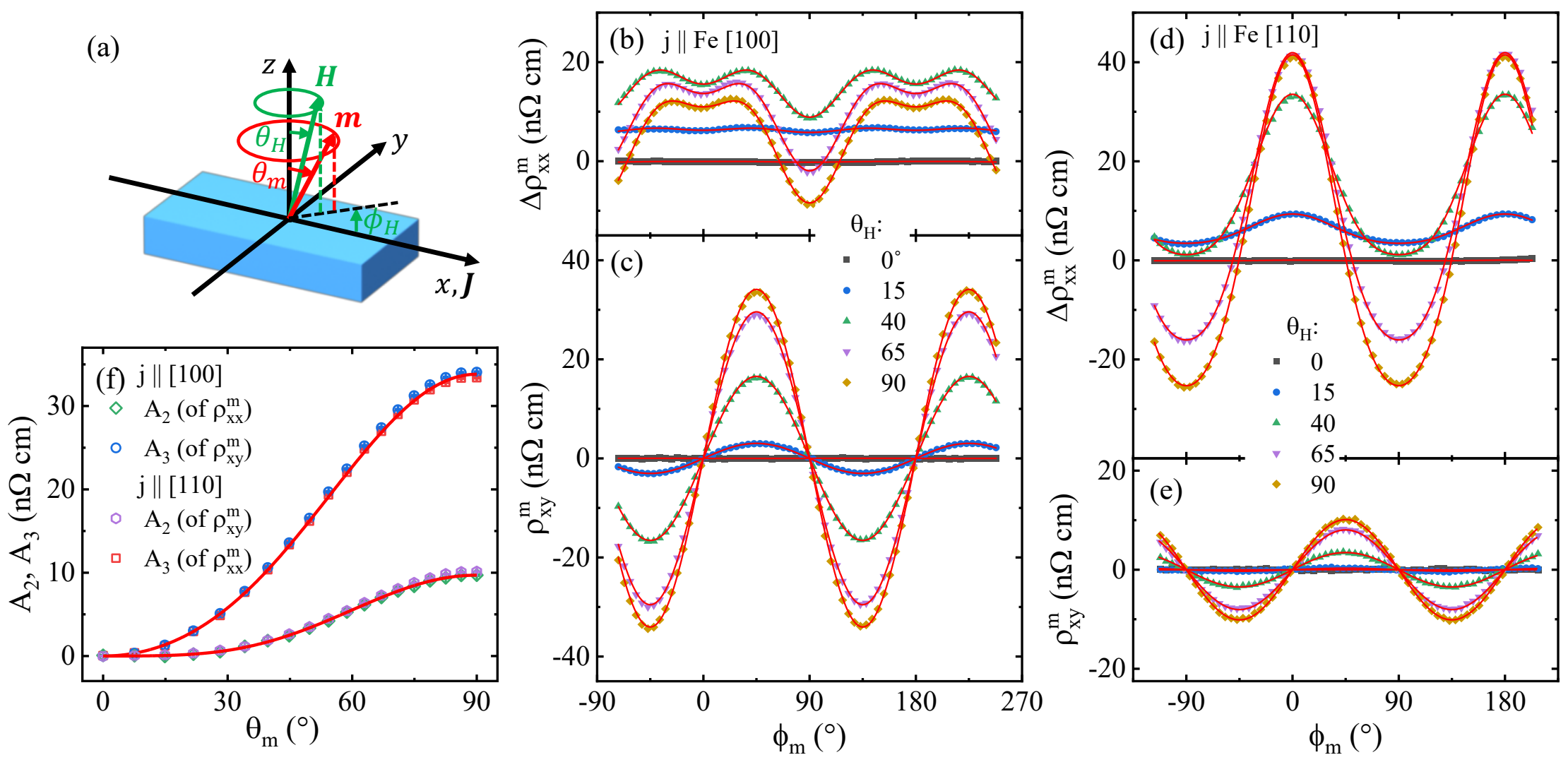

**Fig. 4.** (a) Schematics for measuring AMR and PHE with the field applied across the entire angular space. (b-c) The $\phi_m$-dependence of $\Delta\rho_{xx}^{m}$ and $\rho_{xy}^{m}$ in the device with $\boldsymbol{J} \parallel [100]$. (d-e) The $\phi_m$-dependence of $\Delta\rho_{xx}^{m}$ and $\rho_{xy}^{m}$ in the device with $\boldsymbol{J} \parallel [110]$. The field-induced magnetoresistance and related PHE signals have been subtracted to isolate the magnetization effects. The red lines in (b-e) are fitting curves using Eq. (6). (f) The $\theta_m$-dependence of $A_2$ and $A_3$ in both devices. The lines in (f) represent the fitting curves based on Eq. (4).

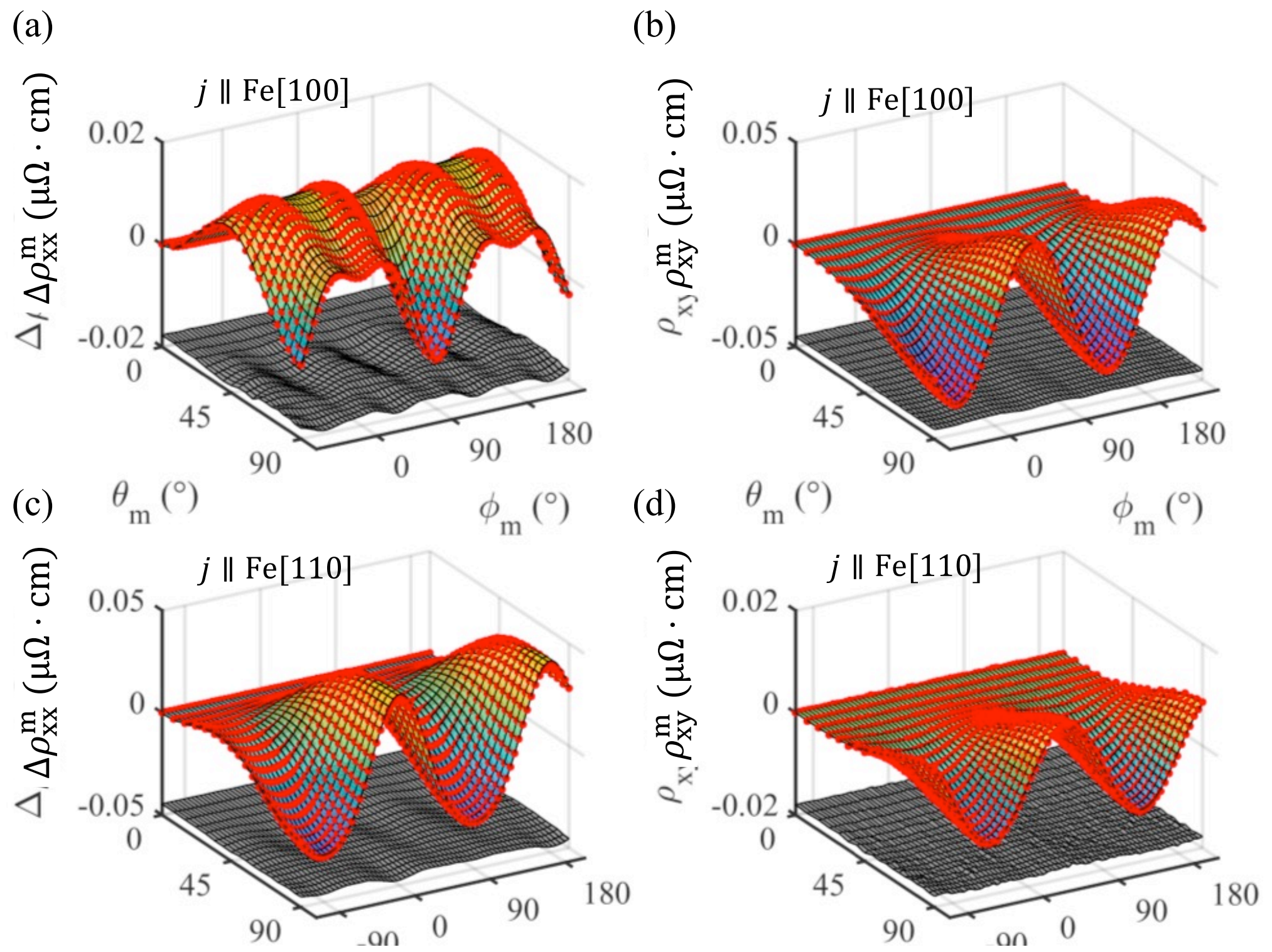

**Fig. 5.** The $\theta_{\boldsymbol{m}}$- and $\phi_{\boldsymbol{m}}$-dependences of $\Delta\rho_{xx}^{\boldsymbol{m}}$ and $\rho_{xy}^{\boldsymbol{m}}$ in the devices with (a-b) $\boldsymbol{J} \parallel [100]$ and (c-d) $\boldsymbol{J} \parallel [110]$ in 3-dimensional plots. Experimental data are shown as red dots. In each figure, the upper surface corresponds to the fitting of Eq. (6), and the lower surface depicts the discrepancy between the experimental data and the fitting results.

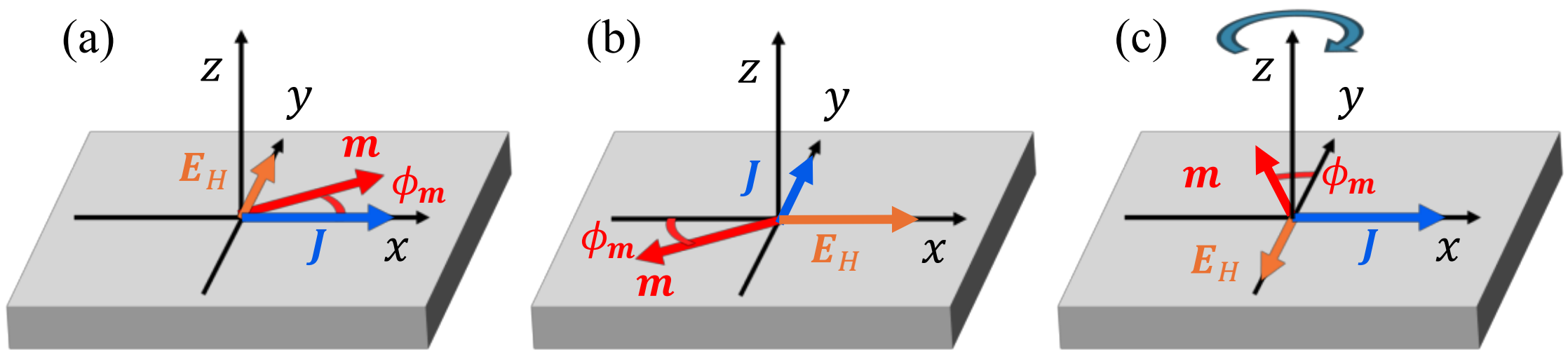

**Fig. 6.** (a) The measurement geometry for PHE in a thin film. (b) The geometry after the swapping of $\boldsymbol{J}$ and $\boldsymbol{E}_{\perp}$, and the inversion of $\boldsymbol{m}$, in accordance with Onsager reciprocal relation. (c) The geometry obtained by applying a 90° counterclockwise rotation about the z-axis to the system depicted in (b).

**Table I.** Fitting parameters for Eq. (6). Experimental results measured under magnetic fields of 6 T and 4 T were fitted. The units for $\rho_1$ to $\rho_7$ are $n\Omega \cdot cm$.

| $\mu_0 H$ (T) | $\rho_1$ | $\rho_2$ | $\rho_3$ | $\rho_4$ |
|---|---|---|---|---|
| 6 | $-2.02 \pm 0.03$ | $-6.34 \pm 0.05$ | $4.91 \pm 0.08$ | $17.63 \pm 0.16$ |
| 4 | $-2.12 \pm 0.07$ | $-6.34 \pm 0.06$ | $4.88 \pm 0.07$ | $17.60 \pm 0.07$ |
| $\mu_0 H$ (T) | $\rho_5$ | $\rho_6$ | $\rho_7$ | |
| 6 | $1.24 \pm 0.05$ | $1.80 \pm 0.05$ | $-0.58 \pm 0.02$ | |
| 4 | $1.13 \pm 0.08$ | $1.81 \pm 0.07$ | $-0.57 \pm 0.01$ | |

## Supplemental Materials for

# Anisotropic galvanomagnetic effects in single-crystal Fe(001) films elucidated by a phenomenological theory

Haoran Chen[1], Zhen Cheng[1], Yizi Feng[1], Hongyue Xu[1], Tong Wu[1], Chuanhang Chen[1], Yue Chen[2,3,4], Zhe Yuan[3,4], Yizheng Wu[1,5,6*]

[1]*Department of Physics, Applied Surface Physics State Key Laboratory, Fudan University, Shanghai 200433, China*

[2]*Center for Advanced Quantum Studies and Department of Physics, Beijing Normal University, Beijing 100875, China*

[3]*Institute for Nanoelectronic Devices and Quantum Computing, Fudan University, Shanghai 200433, China*

[4]*Interdisciplinary Center for Theoretical Physics and Information Science, Fudan University, Shanghai 200433, China*

[5]*Shanghai Research Center for Quantum Sciences, Shanghai 201315, China*

[6]*Shanghai Key Laboratory of Metasurfaces for Light Manipulation, Fudan University, Shanghai 200433, China*

## I. Phenomenological theory with $\mathrm{C_{4v}}$ symmetry

In the main text, we have detailed the magnetization dependence of the resistivity tensor $\rho_{ij}(m)$ using a MacLaurin series expansion [1-3], which includes only odd-ordered terms:

$$\rho_{ij}(\mathbf{m}) = a_{ij} + a_{klij}m_k m_l + a_{klmnij}m_k m_l m_m m_n.$$

All subscripts $k$, $l$, $m$, and $n$ take the values of 1, 2, and 3 following Einstein summation convention. In the expansion, terms differing only in the permutation of subscripts, such as $a_{klij}m_k m_l$ and $a_{lkij}m_l m_k$, are combined into a single term, like $(a_{klij} + a_{lkij})m_k m_l$. To eliminate redundant degrees of freedom, we use a single parameter to describe terms differing only by permutation. Due to the Onsager reciprocal relation $\rho_{ij}(\boldsymbol{m}) = \rho_{ji}(-\boldsymbol{m})$, we have the constraints of $a_{ij} = a_{ji}$, $a_{klij} = a_{klji}$, and $a_{klmnij} = a_{klmnji}$, reducing the number of free parameters in the expansion. According to Neumann's principle, the resistivity tensor $\rho_{ij}(\mathbf{m})$ should be invariant under symmetric operations of the crystal, further reducing the free parameters.

In our study, the Fe(001) film expitaxied on $\mathrm{MgAl_2O_4}(001)$ substrate exhibits $C_{4v}$ symmetry. The symmetry of the $C_{4v}$ point group includes a four-fold rotation axis [001] and four vertical reflection planes $(100)$, $(110)$, $(010)$, and $(\bar{1}10)$. The transformation matrix for each symmetric operation is listed below:

| | | | | | | | |
|---|---|---|---|---|---|---|---|
| $e$ | $\begin{pmatrix}1&0&0\\0&1&0\\0&0&1\end{pmatrix}$ | $c_{4z}^1$ | $\begin{pmatrix}0&-1&0\\1&0&0\\0&0&1\end{pmatrix}$ | $c_{4z}^2$ | $\begin{pmatrix}-1&0&0\\0&-1&0\\0&0&1\end{pmatrix}$ | $c_{4z}^3$ | $\begin{pmatrix}0&1&0\\-1&0&0\\0&0&1\end{pmatrix}$ |
| $\sigma_v^{0^\circ}$ | $\begin{pmatrix}-1&0&0\\0&1&0\\0&0&1\end{pmatrix}$ | $\sigma_v^{45^\circ}$ | $\begin{pmatrix}0&-1&0\\-1&0&0\\0&0&1\end{pmatrix}$ | $\sigma_v^{90^\circ}$ | $\begin{pmatrix}1&0&0\\0&-1&0\\0&0&1\end{pmatrix}$ | $\sigma_v^{135^\circ}$ | $\begin{pmatrix}0&1&0\\1&0&0\\0&0&1\end{pmatrix}$ |

Only symmetric operations $c_{4z}^1$ ($90^\circ$ rotation around [001] axis) and $\sigma_v^{0^\circ}$ (mirror reflection against (100) plane) need to be considered in tensor reduction, as other symmetric operations are just combinations of $c_{4z}^1$ and $\sigma_v^{0^\circ}$. For example, $c_{4z}^3 = c_{4z}^1 c_{4z}^1 c_{4z}^1$ and $\sigma_v^{45^\circ} = c_{4z}^1 \sigma_v^{0^\circ}$. For the Fe (001) film, the current is confined to the (001) plane, thus only $\rho_{11}$, $\rho_{12}$, $\rho_{21}$, and $\rho_{22}$ have contribution to longitudinal and Hall voltages. Considering the invariance of $\rho_{11}$, $\rho_{12}$, and $\rho_{22}$ under $c_{4z}^1$ operation, we have the following constraints on expansion parameters:

| | | | |
|---|---|---|---|
| $a_{11} = a_{22}$ | $a_{12} = 0$ | | |
| $a_{1111} = a_{2222}$ | $a_{1122} = a_{2211}$ | $a_{3311} = a_{3322}$ | $a_{1112} = -a_{2212}$ |
| $a_{1211} = -a_{1222}$ | $a_{1311} = 0$ | $a_{1322} = 0$ | $a_{2311} = 0$ |
| $a_{2322} = 0$ | $a_{1312} = 0$ | $a_{2312} = 0$ | $a_{3312} = 0$ |
| $a_{111111} = a_{222222}$ | $a_{111122} = a_{222211}$ | $a_{333311} = a_{333322}$ | $a_{112211} = a_{112222}$ |
| $a_{113311} = a_{223322}$ | $a_{113322} = a_{223311}$ | $a_{122212} = a_{111212}$ | $a_{122211} = -a_{111222}$ |
| $a_{111211} = -a_{122222}$ | $a_{123311} = -a_{123322}$ | $a_{113312} = -a_{223312}$ | $a_{111112} = -a_{222212}$ |
| $a_{111311} = 0$ | $a_{111312} = 0$ | $a_{111322} = 0$ | $a_{112212} = 0$ |
| $a_{112311} = 0$ | $a_{112312} = 0$ | $a_{112322} = 0$ | $a_{122311} = 0$ |
| $a_{122312} = 0$ | $a_{122322} = 0$ | $a_{133311} = 0$ | $a_{133312} = 0$ |
| $a_{133322} = 0$ | $a_{222311} = 0$ | $a_{222312} = 0$ | $a_{222322} = 0$ |
| $a_{233311} = 0$ | $a_{233312} = 0$ | $a_{233322} = 0$ | $a_{333312} = 0$ |

If the invariance of $\rho_{11}$, $\rho_{12}$, and $\rho_{22}$ under $\sigma_v^{0^\circ}$ operation is also considered, we have additional constraints:

| | | | |
|---|---|---|---|
| $a_{1112} = 0$ | $a_{1211} = 0$ | $a_{1222} = 0$ | $a_{2212} = 0$ |
| $a_{111112} = 0$ | $a_{111211} = 0$ | $a_{111222} = 0$ | $a_{113312} = 0$ |
| $a_{122211} = 0$ | $a_{122222} = 0$ | $a_{123311} = 0$ | $a_{123322} = 0$ |
| $a_{222212}=0$ | $a_{223312}=0$ | | |

Utilizing these constraints, the $a_{ij}$ , $a_{ijkl}$ , $a_{ijklmn}$ parameters appearing in components $\rho_{11}$, $\rho_{12}$, $\rho_{21}$, and $\rho_{22}$ can be reduced to 8 independent parameters:

$$\rho_0 = a_{11} + \frac{1}{4}(a_{1111} + a_{2211} + 2a_{3311})$$
$$+\frac{3}{64}(3a_{111111} + 3a_{222211} + 6a_{112211} + 8a_{113311} + 8a_{113322} + 8a_{333311})$$
$$\rho_1 = \frac{1}{4}(-a_{1111} - a_{2211} + 2a_{3311}) + \frac{1}{16}(-3a_{111111} - 3a_{222211} - 6a_{112211} + 8a_{333311})$$
$$\rho_2 = \frac{1}{64}(3a_{111111} + 3a_{222211} + 6a_{112211} - 24a_{113311} - 24a_{113322} + 8a_{333311})$$
$$\rho_3 = \frac{1}{4}(a_{1111} - a_{2211}) + \frac{1}{4}(a_{111111} - a_{222211})$$
$$\rho_4 = \frac{1}{2}a_{1212} + a_{111212}$$
$$\rho_5 = \frac{1}{16}(a_{111111} - a_{222211} - 6a_{113311} + 6a_{113322})$$
$$\rho_6 = \frac{1}{4}(a_{111212} - 3a_{123312})$$
$$\rho_7 = \frac{1}{64}(a_{111111} + a_{222211} - 6a_{112211})$$

By introducing new parameters $\rho_0, \rho_1, \ldots, \rho_7$ and express the magnetization unit $\boldsymbol{m}$ in spherical coordinates with $\boldsymbol{m} = (\sin\theta_{\mathbf{mc}}\cos\phi_{\mathbf{mc}}, \sin\theta_{\mathbf{mc}}\sin\phi_{\mathbf{mc}}, \cos\theta_{\mathbf{mc}})$ , the resistivity tensor can be represented in a more concise form:

$$\begin{aligned}
\rho_{11}(\theta_{\mathbf{mc}}, \phi_{\mathbf{mc}}) &= A_1 + A_2\cos 2\phi_{\mathbf{mc}} + A_4\cos 4\phi_{\mathbf{mc}}, \\
\rho_{22}(\theta_{\mathbf{mc}}, \phi_{\mathbf{mc}}) &= A_1 - A_2\cos 2\phi_{\mathbf{mc}} + A_4\cos 4\phi_{\mathbf{mc}}, \\
\rho_{12}(\theta_{\mathbf{mc}}, \phi_{\mathbf{mc}}) &= \rho_{21}(\theta_{\mathbf{mc}}, \phi_{\mathbf{mc}}) = A_3\sin 2\phi_{\mathbf{mc}}.
\end{aligned}$$

$A_1$, $A_2, A_3$ and $A_4$ are the $\theta_{\mathbf{mc}}$-dependent coefficients, which can be expressed as

$$\begin{aligned}
A_1(\theta_{\mathbf{mc}}) &= \rho_0 + \rho_1\cos 2\theta_{\mathbf{mc}} + \rho_2\cos 4\theta_{\mathbf{mc}}, \\
A_2(\theta_{\mathbf{mc}}) &= \rho_3(1 - \cos 2\theta_{\mathbf{mc}}) - \rho_5(1 - \cos 4\theta_{\mathbf{mc}}), \\
A_3(\theta_{\mathbf{mc}}) &= \rho_4(1 - \cos 2\theta_{\mathbf{mc}}) - \rho_6(1 - \cos 4\theta_{\mathbf{mc}}), \\
A_4(\theta_{\mathbf{mc}}) &= \rho_7(3 - 4\cos 2\theta_{\mathbf{mc}} + \cos 4\theta_{\mathbf{mc}}).
\end{aligned}$$

These equations correspond to Eqs. (3) and (4) in the main text.

## II. Phenomenological theory with $\mathrm{O_h}$ symmetry

We have also derived expressions for $\rho_{11}$, $\rho_{12}$, $\rho_{21}$, and $\rho_{22}$, considering $O_h$ symmetry. Compared to $C_{4v}$ symmetry, $O_h$ symmetry introduces an additional four-fold rotation symmetry around [100] axes. The higher symmetry imposes more constraints on the expansion coefficients, which are $a_{2211} = a_{3311}$, $a_{112211} = a_{113311}$, and $a_{222211} = a_{333311}$. Consequently, the expressions for the parameters $\rho_0, \rho_1, \dots, \rho_7$ are modified as follows:

$$\rho_0 = a_{11} + \frac{1}{4}(a_{1111} + 3a_{2211}) + \frac{3}{64}(3a_{111111} + 11a_{222211} + 14a_{112211} + 8a_{113322})$$

$$\rho_1 = \frac{1}{4}(-a_{1111} + a_{2211}) + \frac{1}{16}(-3a_{111111} + 5a_{222211} - 6a_{112211})$$

$$\rho_2 = \frac{1}{64}(3a_{111111} + 11a_{222211} - 18a_{112211} - 24a_{113322})$$

$$\rho_3 = \frac{1}{4}(a_{1111} - a_{2211}) + \frac{1}{4}(a_{111111} - a_{222211})$$

$$\rho_4 = \frac{1}{2}a_{1212} + a_{111212}$$

$$\rho_5 = \frac{1}{16}(a_{111111} - a_{222211} - 6a_{112211} + 6a_{113322})$$

$$\rho_6 = \frac{1}{4}(a_{111212} - 3a_{123312})$$

$$\rho_7 = \frac{1}{64}(a_{111111} + a_{222211} - 6a_{112211})$$

Under $O_h$ symmetry, the number of independent parameters is reduced from 8 in $C_{4v}$ symmetry to 6, due to the additional constraints. These constraints require $\rho_i$ parameters to conform to $\rho_1 - 4\rho_2 + \rho_3 - 4\rho_5 + 24\rho_7 = 0$ and $\rho_1 + \rho_3 - 4\rho_7 = 0$.

This result is consistent with previous theory of galvanomagnetic effects based on $O_h$ symmetry [3,4]. However, a notable difference compared to the theory of Miao et al. [5] is the absence of four-fold planar Hall effect in our theory.

## III. Determination of field-induced planar Hall effect

In the main text, we demonstrated that the field-induced magnetoresistance (FIMR) significantly contributes to the planar Hall resistivity $\rho_{xy}$. The parameters $R_{xy,0}^{\mathrm{MMR}}$ and $R_{xy,0}^{\mathrm{OMR}}$ in our model for FIMR are determined by sweeping the magnetic field $45^\circ$ and 135° away from the current within the film plane. Fig. S2 (a) shows the field dependence of $\rho_{xy}$ in the J // [110] device, along 0°, 45°, 90°, and 135° directions within the film plane. In the field dependences of $\rho_{xy}$ with 0° and 90° field orientations, the curves exhibit little variation, indicating that the field-induced $\rho_{xy}^{\mathrm{FIMR}}$ is neglectable. In the cases of 45° and 135° field orientations, $\rho_{xy}$ exhibits a significant field-dependent variation, and the field-induced $\rho_{xy}^{\mathrm{FIMR}}$ has opposite sign along these two directions. Compared with the magnetization-induced $\rho_{xy}$ at zero field, the field-induced $\rho_{xy}^{\mathrm{FIMR}}$ at 6 T is even larger. Consequently, it is imperative to eliminate the FIMR contribution in order to extract pure $m$-induced $\rho_{xy}^{m}$. To determine the FIMR contribution $\rho_{xy}^{\mathrm{FIMR}}$, we utilized Eq. (8) in the main text to fit the curves measured along 45°, and obtain $R_{xy,0}^{\mathrm{MMR}}$ and $R_{xy,0}^{\mathrm{OMR}}$ as the coefficient of linear and quadratic terms, respectively.

Fig. S2 (b) shows the field dependence of $\rho_{xy}$ in the J//[100] device. The field-induced $\rho_{xy}^{\mathrm{FIMR}}$ is relatively small compared with the $m$-induced $\rho_{xy}^{m}$ at zero field. But we still obtain $R_{xy,0}^{\mathrm{MMR}}$ and $R_{xy,0}^{\mathrm{OMR}}$ for J//[100] device by fitting the curve measured at 45° and eliminate the FIMR contribution of $\rho_{xy}$.

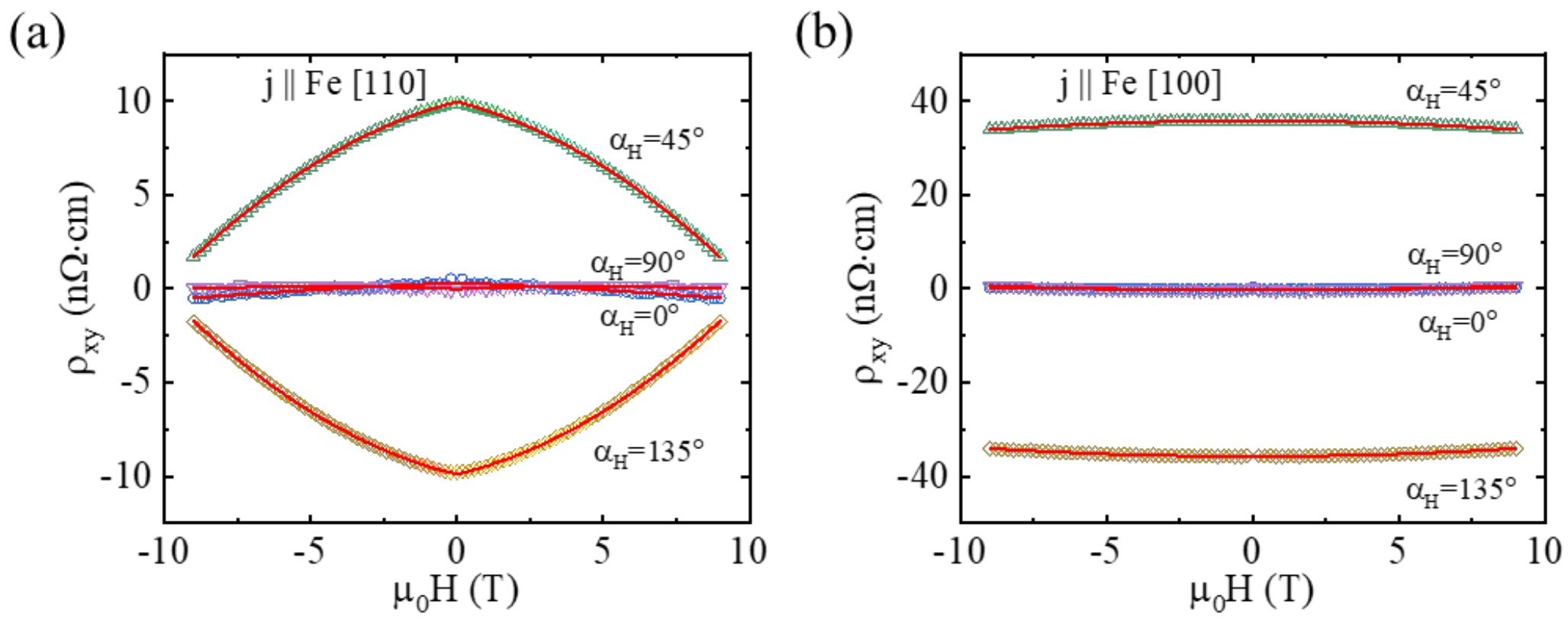

**Fig. S1** The field dependence of $\rho_{xy}$ in (a) J//[110] and (b) J//[100] devices along 0°, 45°, 90°, and 135° within the film plane.

## IV. Elimination of field-induced magnetoresistance with field rotating in the xz plane

In the main text, the $\alpha_H$-dependences of $\rho_{xx}$ and $\rho_{xy}$ measured with the rotation of 6 T and 4 T magnetic field are found to be different, due to the field-induced magnetoresistance (FIMR). We successfully established an anisotropic model for FIMR and determined the FIMR contribution to longitudinal resistivity $\rho_{xx}^{\mathrm{FIMR}}(\boldsymbol{H})$ and Hall resistivity $\rho_{xy}^{\mathrm{FIMR}}(\boldsymbol{H})$ with arbitrary field orientation. Upon removing the FIMR contributions, the $\alpha_m$ -dependences of magnetization-induced $\rho_{xx}^{m}$ and $\rho_{xy}^{m}$ measured with 6 T and 4 T field are found to be identical.

We also compared the $\gamma_H$-dependences of $\rho_{xx}$ measured with 6 T and 4 T field, represented by blue and red dots, respectively, in Fig. S3 (a). Upon aligning the 4 T curve with the 6 T curve, it becomes evident that the differences between the two are not limited to their offsets but also extend to the shape of the curves.

After eliminating the FIMR contribution $\rho_{xx}^{\mathrm{FIMR}}$ as detailed in the main text, we present the $\gamma_m$-dependence of $\rho_{xx}^{m}$ in Fig. S3 (b). The resulting curves for the rotating fields of 6 T and 4 T are identical, confirming that our method for deducing the field-induced contributions in $\rho_{xx}$ is also effective within the xz plane.

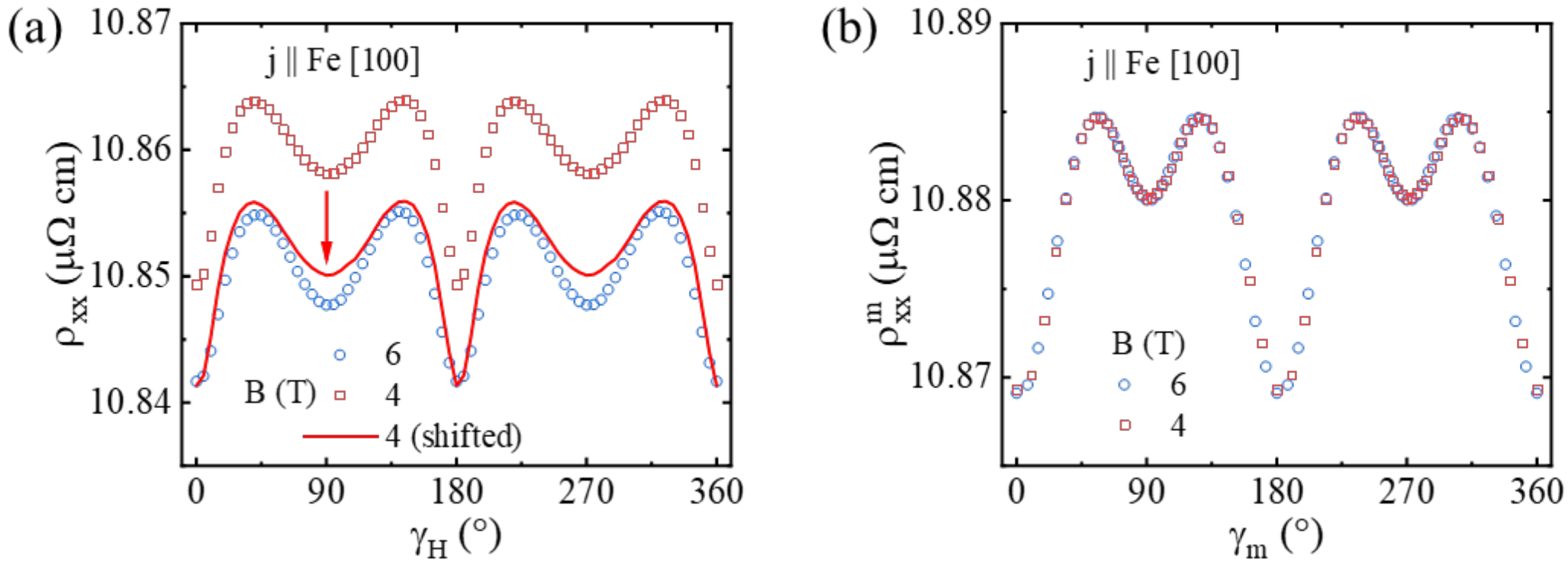

**Fig. S2** (a) The $\gamma_m$-dependence of $\rho_{xx}$ measured with different field strengths. The red line is the shift $\rho_{xx}$ curve of 4 T to better emphasize the anisotropy of the field-induced signals $\rho_{xx}^{FIMR}$. (b) The intrinsic magnetization-induced $\rho_{xx}^{\boldsymbol{m}}$ with the field-induced signals $\rho_{xx}^{\mathrm{FIMR}}$ subtracted to isolate the magnetization effect.

## V. Full angular mapping of $\rho_{xx}(m)$ and $\rho_{xy}(m)$ with 6 T and 4 T field

In the main text, we conducted full angular-mapping measurements of the magnetization-induced longitudinal resistivity $\rho_{xx}^{m}$ and planar Hall resistivity $\rho_{xy}^{m}$, using applied fields of 6 T and 4 T, and fitted these data to our phenomenological theory. Field-induced magnetoresistance (FIMR) has been subtracted both from $\rho_{xx}(m)$ and $\rho_{xy}(m)$, making the fitting results presented in Table I of the main text independent of the applied magnetic field strength.

Additionally, we present the raw experimental data without FIMR subtraction in Fig. S4, which shows the similar angular dependence of intrinsic galvanomagnetic signal in Fig. 5 in the main text. The fitting results for both the raw data and the data with FIMR eliminated are detailed in Table S1. In our angular-dependent model for FIMR, only two-fold terms are considered. Thus, eliminating FIMR will make correction on the two-fold AMR parameters $\rho_1$, $\rho_3$, and $\rho_4$. In the fitting results of raw data, $\rho_3$ shows a deviation of 34% between 6 T and 4 T measurements. However, upon FIMR removal, this deviation is significantly reduced to 1%. The deviation of $\rho_1$ between 6 T and 4 T measurements also decreases from 18% to 5% after FIMR is eliminated. $\rho_4$ is not sensitive to the strength of applied field, because FIMR has little anisotropy with $J \parallel [110]$, as illustrated in Fig. 1(c) in the main text. For all fitting parameters, the deviation between 6 T and 4 T drops below the error of fitting parameters after the subtraction of FIMR. We successfully obtained intrinsic magnetization-induced magnetoresistance, which is independent of the strength of magnetic field.

Table SI. Fitting results of $\rho_{xx}(\theta_m, \phi_m)$ and $\rho_{xy}(\theta_m, \phi_m)$ with phenomenological theory. The fitting results from both raw data and data with FIMR eliminated are listed.

| (μΩ · cm) | Raw data | | | Data with FIMR eliminated | | |
|---|---|---|---|---|---|---|
| | 6 T | 4 T | Mismatch | 6 T | 4 T | Mismatch |
| $\rho_1$ | $-2.74 \pm 0.03$ | $-3.22 \pm 0.04$ | 18% | $-2.02 \pm 0.03$ | $-2.12 \pm 0.07$ | 5% |
| $\rho_2$ | $-6.22 \pm 0.05$ | $-6.05 \pm 0.03$ | 3% | -6.34 $\pm$ 0.05 | -6.34$\pm$0.06 | 0.0% |
| $\rho_3$ | $2.70 \pm 0.07$ | $3.63 \pm 0.08$ | 34% | $4.91 \pm 0.08$ | $4.88 \pm 0.07$ | 1% |
| $\rho_4$ | $17.27 \pm 0.15$ | $17.42 \pm 0.10$ | 1% | $17.63 \pm 0.16$ | $17.60 \pm 0.07$ | 0.2% |
| $\rho_5$ | $0.87 \pm 0.05$ | $0.82 \pm 0.07$ | 6% | $1.24 \pm 0.05$ | $1.13 \pm 0.08$ | 9% |
| $\rho_6$ | $1.74 \pm 0.01$ | $1.77 \pm 0.08$ | 2% | $1.80 \pm 0.05$ | $1.81 \pm 0.07$ | 0.6% |
| $\rho_7$ | $-0.58 \pm 0.02$ | $-0.57 \pm 0.01$ | 2% | $-0.58 \pm 0.02$ | $-0.57 \pm 0.01$ | 2% |

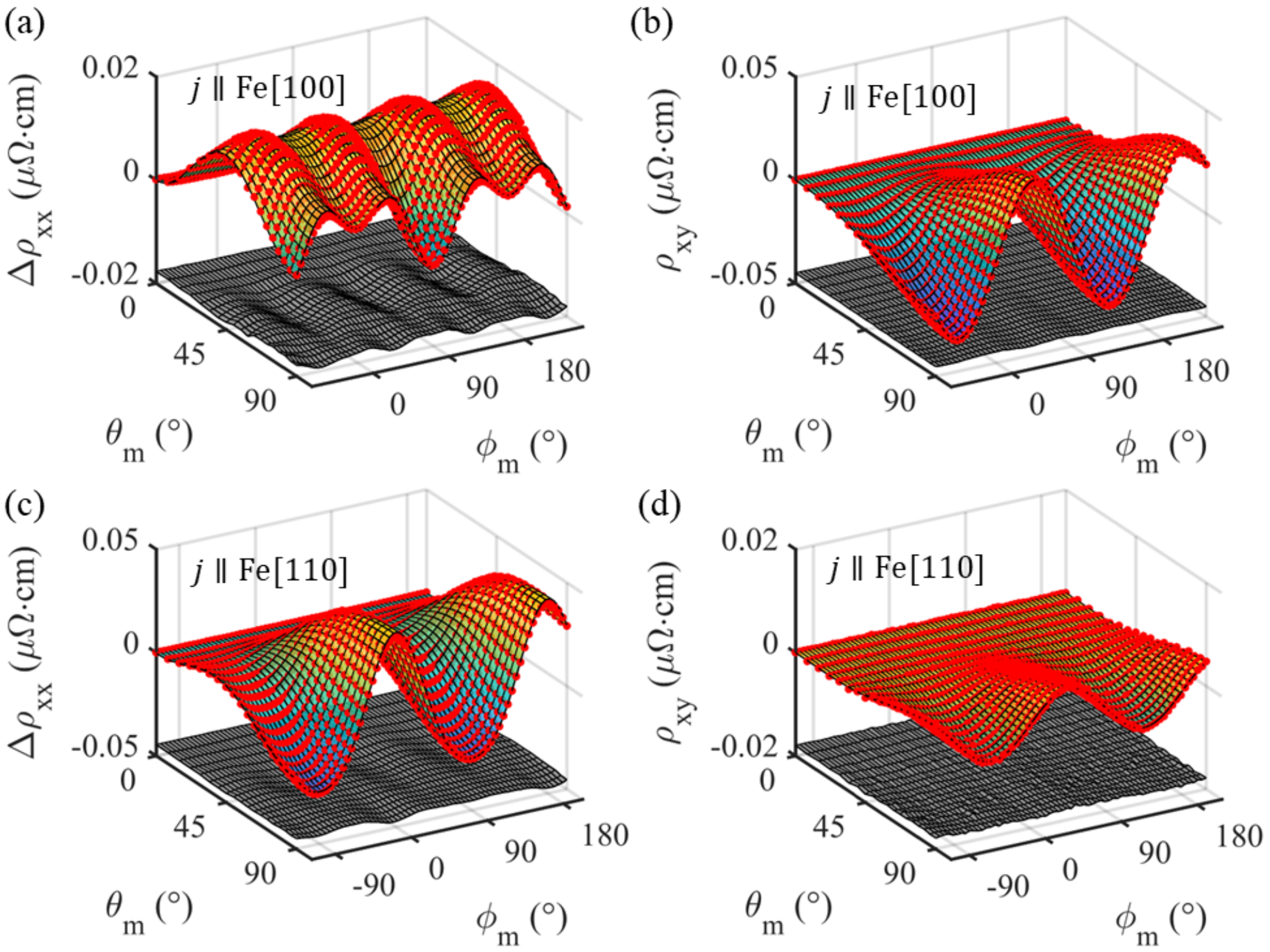

**Fig. S3** The $\theta_m$- and $\phi_m$-dependences of $\Delta\rho_{xx}^{m}$ and $\rho_{xy}^{m}$ in the devices with (a-b) $\boldsymbol{J} \parallel [100]$ and (c-d) $\boldsymbol{J} \parallel [110]$ in 3-dimensional plots. Experimental data with FIMR included are shown as red dots. In each figure, the upper surface corresponds to the fitting of Eq. (6), and the lower surface depicts the discrepancy between the experimental data and the fitting results.